\documentclass[]{aa}
\usepackage{graphicx,natbib}
\usepackage{txfonts}
\begin{document}

   \title{Properties of bow-shock sources at the Galactic center}
   \author{J. Sanchez-Bermudez \inst{\ref{inst1},\ref{inst2}} \and
     R. Sch\"odel \inst{\ref{inst1}} \and A. Alberdi
     \inst{\ref{inst1}}\and K. Mu$\breve{z}$i\'c \inst{\ref{inst3}}
     \and C. A. Hummel \inst{\ref{inst2}} \and J.-U. Pott \inst{\ref{inst4}}}
      \institute{Instituto de Astrof\'isica de Andaluc\'ia (CSIC), Glorieta de la Astronom\'ia S/N, 18008 Granada, Spain. 
              \label{inst1} \email{joel@iaa.es}
         \and
        European Southern Observatory, Karl-Schwarzschild-Stra$\beta$e 2, 85748 Garching, Germany.
             \label{inst2}
             \and
             European Southern Observatory, Alonso de C\'ordova 3107,
             Casilla 19, Santiago, Chile.\label{inst3}
         \and
        Max-Planck-Institut f\"ur Astronomie, K\"onigstuhl 17, D-69117 Heidelberg, Germany.
             \label{inst4}
}    
 \abstract
   {There exists an enigmatic population of massive stars around
     the Galactic center (GC) that were formed some Myr ago. A fraction of these stars has been found to orbit the
     supermassive black hole, Sgr\,A*, in a projected clockwise disk-like structure, which suggests that they were formed in a
     formerly existing dense disk around Sgr\,A*.
}
   { We focus on a subgroup of objects, the extended,
     near-infrared (NIR) bright sources IRS\,1W, IRS\,5, IRS\,10W, and
     IRS\,21, that have been suggested to be young, massive
     stars that form bow shocks through their interaction with the interstellar medium (ISM). Their nature has impeded
     accurate determinations of their orbital parameters. We aim at
     establishing their nature and kinematics to test whether
     they form part of the clockwise disk.}
   {We performed NIR multiwavelength
     imaging with NACO/VLT using direct adaptive optics (AO) and AO-assisted sparse aperture masking (SAM). We introduce a
     new method for self-calibration of the SAM  point spread
function
     in dense stellar fields. The emission mechanism, morphology, and kinematics
     of the targets were examined via 3D models, combined with existing models of the gas flow in the
     central parsec.
}
   {We confirm previous findings  that IRS\,21, IRS\,1W, and IRS\,5 are bow-shocks created by the interaction between mass-losing
     stars and the interstellar gas. The nature of IRS\,10W remains unclear. Our modeling shows that the
     bow-shock emission is caused by thermal emission, while the
     scattering of stellar light does not play a
     significant role.  IRS\,1W shows a morphology that is consistent with a
     bow shock produced by an anisotropic stellar wind or by locally
     inhomogeneous ISM density. Our best-fit models provide estimates of the local
     proper motion of the ISM in the Northern Arm that agree with previously published models that were based on radio interferometry and NIR spectroscopy. Assuming that all of the sources are gravitationally tied to Sagittarius A*, their orbital planes were obtained via a Monte Carlo simulation.}
   {Our sources appear to be Wolf-Rayet stars associated to
     the last starburst at the GC. Our orbital
     analysis suggests that they are not part of any
     of the previously suggested coherent stellar structures, in
     particular the clockwise disk. We thus add more evidence to
     recent findings that a large proportion of the massive stars show apparently random orbital orientations, suggesting either that not all of them were formed in the clockwise disk, or that their orbits were randomized rapidly after formation in the disk.}

   \titlerunning{Bow shocks in the Galactic center}

   \maketitle
%::::::::::::::::::::::::
\section{Introduction}

Sagittarius\,A\,West, the so-called mini-spiral, is a prominent
feature of the interstellar medium (ISM) in the inner $\sim$2\,pc of
the Milky Way \citep{Ekers:1983uq,Lo:1983kx}. It is made up of
three streamers of ionized gas, termed the Northern, Western, and
Eastern arms. Near- and mid-infrared observations of the mini-spiral
show that the ionized gas is mixed with warm and hot dust that is in
part arranged in complex structures like: extended filaments, bow-shocks and compact, barely resolved knots
\citep[][]{Clenet:2004ve,Viehmann:2006uq,Muzic:2007kx}.  Studies of
the radial velocities and proper motions of the gas in the mini-spiral
have shown that their dynamics can generally be
described by three bundles of quasi-Keplerian orbits in the combined
potential of the central stellar cluster of the Milky
Way plus a
  supermassive black hole
\citep[e.g.,][]{Vollmer:2000vn,Paumard:2004kx,Zhao:2009la}.

It is generally thought that the gas in the mini-spiral is ionized by
the radiation of more than one hundred massive, young stars that are
located mainly within 0.5\,pc of the supermassive black hole,
Sagittarius\,A*, and that were formed in a star formation event about
2-6\,Myr ago \citep{Paumard:2006xd,Bartko:2010fk, Lu_2013}. Some of these
young stars appear to orbit the black hole in a disk-like structure
that rotates in a clockwise direction (referred to as clockwise
system, CWS, in the following). This observation, along with the
finding that the young stars are concentrated toward the black hole,
gave rise to the currently favored hypothesis for their formation {\it
  in situ} in a formerly existing gas disk around Sagittarius A*
\citep{Levin:2003kx,Paumard:2006xd,Lu:2009bl,Bartko:2009fq}. Considerable
open questions in this picture remain, however, in particular,
concerning the nature of the young stars that are not part of the CWS
\citep[see, e.g., discussion in][]{Genzel:2010fk}: is the second
postulated system of a counterclockwise streamer or disk (CCWS) real? Is the CWS strongly warped? Can the current kinematical configuration of the young, massive stars be reached when one
assumes that all of them are on the CWS? It is therefore of great interest to obtain as much information as possible about the orbital parameters of all young, massive stars.

Very many bow-shocks show evidence of interactions between
stellar winds from the young massive stars and the ISM in the
mini-spiral
\citep{Clenet:2004ly,Tanner:2005fk,Geballe:2006fk,Muzic:2007kx,Buchholz:2011fk}. Some
of these bow shocks are among the brightest near- to mid-infrared
sources in the central parsec of the Galactic center (GC) and have been
known for decades \citep[][]{Rieke:1973fk,Becklin:1975uq} before
their nature was finally revealed through radiative transfer modeling
and, in particular, high angular resolution imaging
\citep{Tanner:2002kx,Tanner:2005fk}.

\citet{Tanner:2005fk} analyzed high angular resolution near-infrared
(NIR) observations of the sources IRS\,1W, 5, 10W, and 21 and
modeled their bow shocks. From the large measured stand-off distances
of the bow shocks' apices the authors concluded that the central sources must be Wolf-Rayet stars and thus were created in the above-mentioned star formation event. Because of the highly embedded nature of the sources, no spectroscopic confirmation of this classification could be obtained so far, with the exception of a tentative identification of IRS\,1W as a Be star by \citet{Paumard:2006xd}. The lack of suitable spectra means that the line-of-sight velocities of the bow-shock stars are unknown and cannot be used to estimate their orbital parameters.

Nevertheless, bow-shock (BS) sources are of great interest because
they allow us, in principle, to infer all six of their phase space
parameters, and thus one more (the line-of-sight distance) than in the
case of other young stars. Their 2D positions and
proper motions can be inferred from imaging. Their line-of-sight (LOS)
velocities and distances can be inferred from their interaction with
the mini-spiral and from the resulting shape of the bow shocks in
combination with models of the gas flow in the mini-spiral
\citep[e.g.,][]{Vollmer:2000vn,Paumard:2004kx,Zhao:2009la}. In this
way, \citet{Tanner:2005fk} were able to show that the kinematic
properties of IRS\,1W and 10W, as determined from their measurements,
are consistent with the hypothesis that these stars belong to the
CWS. However, they did not have the necessary proper motion
measurements to perform this kind of analysis on IRS\,5. For IRS\,21, \citet{Tanner:2005fk} were unable to
resolve its bow-shock and concluded that it was seen face-on
\citep[see also][]{Tanner:2002kx}. \citet{Buchholz:2011fk}, however,
were recently able to resolve the bow-shock structure of IRS\,21
through deconvolution of high-quality adaptive optics data and could thus discard the face-on hypothesis.

Here we continue the investigation of the bow-shock sources IRS\,1W,
5, 21, and 10W by \citet{Tanner:2005fk}. We use 
our own multiwavelength AO assisted imaging as well as AO-assisted
sparse aperture masking (SAM) \citep[see, e.g.,][]{Tuthill:2006fk} to investigate the structure of the sources, which we then fit with fully 3D models. The results from the modeling are then combined with new, more accurate proper motion measurements and with models of the mini-spiral flow to investigate the kinematics of the bow-shock stars.

Section\,\ref{sec:data} describes the acquisition and reduction of the
data. The calibration process and image
reconstruction of our SAM observations is described in particular
detail because we describe a novel way of self-calibrating SAM data in
a dense stellar field. This procedure can be of general interest for
SAM observations of targets in dense stellar
fields. Section\,\ref{sec:proper} describes the results of the proper
motion measurements. Section\,\ref{sec:BSmodel} describes the
mathematical model used in determining the shape and emission
of the bow shocks and the corresponding fitting procedure. Section
\ref{sec:Orbital_param} describes the local kinematics of the Northern
Arm at the position of the bow shocks and the procedure of determining the probability density
function of the source orbital planes. Finally, in
Sect.\,\ref{sec:discussion} we discuss our results and present our conclusions.

\section{Observations and data reduction \label{sec:data}}

\subsection{AO imaging}

We used AO-assisted imaging observations of the central parsec of the
GC with VLT/NACO to measure the proper motions of the bow-shock
sources and to infer their structure. Table\,\ref{Tab:ImObs}
summarizes the observations. The bright supergiant IRS\,7, located
about $5.5"$ north of Sgr\,A*, was used to close the loop of the
AO. IRS\,7 is a variable source \citep{Ott:1999ly}. The most recently
published values for its flux are $K_{s}=7.69\pm0.06$ in April 2006
and $K_{s}=6.96\pm0.04$ in March 2009
\citep[see][]{Schodel:2012fk}. All imaging data were sky subtracted,
flat fielded, corrected for bad pixels, and then combined via a
shift-and-add algorithm into final mosaics, as described, for
instance, in
\citet{Schodel:2009zr}. The saturated cores of bright stars were
repaired in all images from epochs 2002 to 2008. The images from
2009 to 2011 did not need to be repaired since saturation was not a problem because of the short
exposure times used for these observations. The data from 2010 and 2011 were taken with NACO's {\it cube
  mode,} which registers each individual exposure
\citep[see][]{Girard:2010fk}. To improve the image quality of the
latter data we applied a {\it lucky imaging} method and selected the
$30\%$ of the frames with the highest Strehl ratios.

To remove the seeing halos in the AO images, we extracted point
spread functions (PSFs) from
the $H$- and $K_{s}$- band images that were obtained with the S13
camera of NACO and deconvolved the images  with the
Lucy-Richardson algorithm \citep{Lucy:1974os}.  The deconvolved images
were restored with a Gaussian beam of 2\,pixels FWHM to create the
final images to be used in our analysis.

\begin{table}
\centering
\caption{VLT/NACO imaging observations}
\label{Tab:ImObs} 
\begin{tabular}{lllll}
\hline
\hline
Date & Filter & N$^{\mathrm{a}}$  & DIT$^{\mathrm{b}}$\,[s] & Camera$^{\mathrm{c}}$\\
\hline
03 May 2002$^{\mathrm{d}}$ & $K_{s}$ & 60 & 20  & S27\\
10 May 2003$^{\mathrm{d}}$ & $K_{s}$ & 2280  & 0.5 & S27 \\
12 Jun 2004$^{\mathrm{d}}$ & $IB\_2.06$ & 96 & 30 & S27 \\
13 May 2005$^{\mathrm{d}}$ & $K_{s}$ & 6180 & 0.5 &S27 \\
29 Apr 2006$^{\mathrm{d}}$ & $K_{s}$ & 896 & 2 & S27 \\
28 May 2008$^{\mathrm{d}}$ & $K_{s}$ & 104 & 10 & S27 \\
31 Mar 2009$^{\mathrm{d}}$ & $K_{s}$ & 1920 & 1 & S27 \\
28 Sep 2010$^{\mathrm{d}}$ & $K_{s}$ & 2016 &  1.0 & S27 \\
17 May 2011$^{\mathrm{d}}$ & $K_{s}$ & 143 &  2.0 & S27 \\
\hline
03 May 2009$^{\mathrm{e}}$ & $K_{s}$ & 109 & 3 & S13 \\
21 Jul 2009$^{\mathrm{e}}$ & $H$ & 138 & 10 & S13 \\
03 Jun 2010$^{\mathrm{f}}$ & $L'$ & 20 & 5 & L27 \\
03 Jun 2010$^{\mathrm{f}}$ & $NB\_3.74$ & 6 & 20 & L27 \\
04 Jun 2010$^{\mathrm{f}}$ & $L'$ & 20/30 & 5 & L27 \\
\hline
\end{tabular}
\begin{list}{}{}
\item[$^{\mathrm{a}}$] Number of exposures.
\item[$^{\mathrm{b}}$] Detector integration time. The total integration time of each observation amounts to N$\times$DIT.
\item[$^{\mathrm{c}}$] The pixel scale of the NACO S27 camera is
  $0.027"$/pixel, that of S13 is $0.013"$/pixel .
\item[$^{\mathrm{d}}$] Used for proper motion measurements.
\item[$^{\mathrm{e}}$] Imaging observations used to investigate the structure of the bow shocks.
\item[$^{\mathrm{f}}$] SAM observations used to investigate the structure of the bow shocks.
\end{list}
 \end{table}

\subsection{Observations with sparse aperture masking \label{sec:observatios}}

Our SAM observations of the GC central parsec were performed in the
nights of 3/4 and 4/5 June 2010 with VLT/NACO. We used the L27 camera
($0.027"$/pixel) and chose two combinations of filters and mask: the
$L'$ broad-band filter ($\lambda_{central}=3.80\,\mu$m,
$\Delta\lambda=0.62\,\mu$m) was combined with the \textit{ BB\_9holes}
mask and the $NB\_3.74$ narrow-band filter
($\lambda_{central}=3.74\,\mu$m, $\Delta\lambda=0.02\,\mu$m) with the \textit{9\_holes} mask.  The latter combination
reduces bandwidth smearing, while the first combination leads to
higher sensitivity.  The observations were recorded using pupil
tracking. 

The wide field of view (FOV) ($27''\times27"$) of the L27 camera
allowed us to observe all targets simultaneously. All SAM data were
pre-processed according to standard imaging data reduction, that
is, sky
subtraction, flat-fielding, and bad-pixel correction. 

\begin{figure}
\centering
\includegraphics[width=\columnwidth]{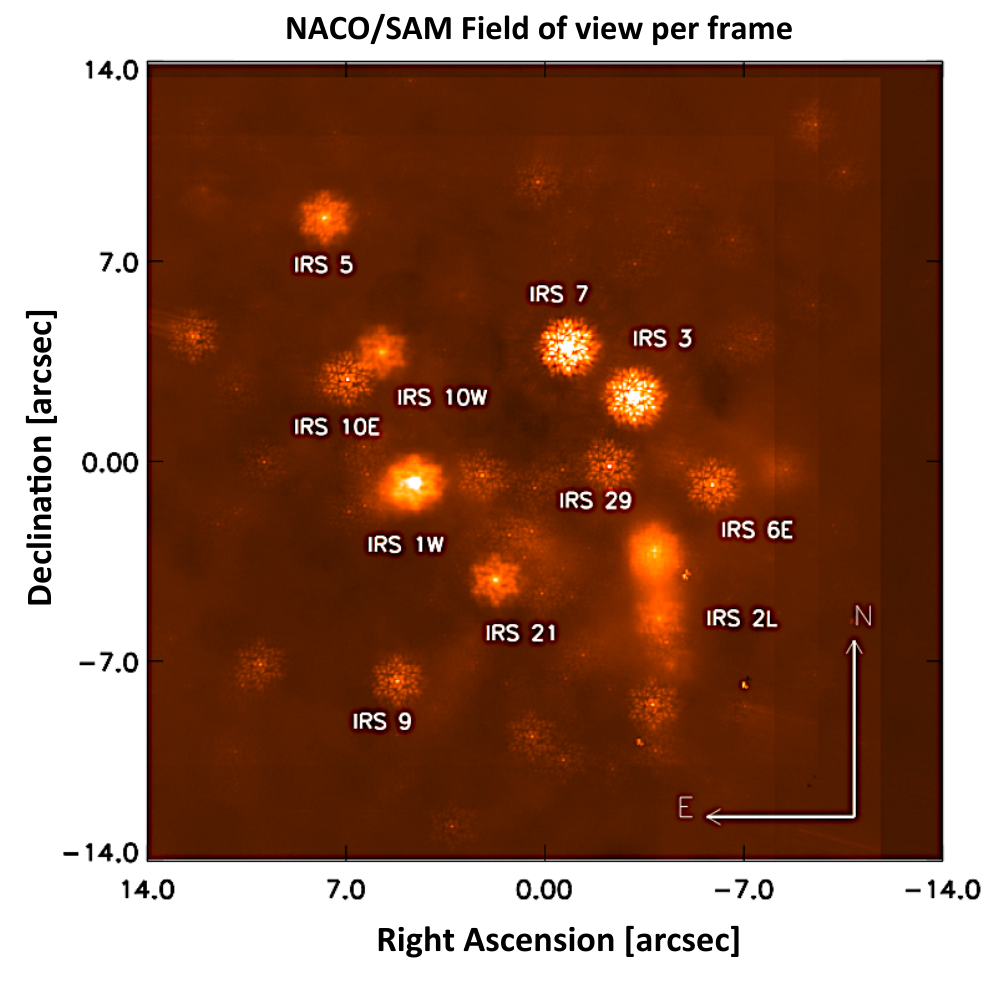} 
\caption{Sparse aperture masking frame of the Galactic
  center. Prominent sources are labeled with their
  names. Positions are given as distances from the center of the
  field of view.}
\label{fig:samim}
\end{figure}

Depending on the observational data sets, the sky was obtained by
calculating either (i) the median of separate sky observations
obtained on a dark cloud about $400"$ north and $700"$ west of
Sagittarius\,A*, or (ii) via the median of the appropriately clipped
values at each pixel from the dithered images on-target when no
dedicated sky observations close in time were available.  We note that
sky subtraction did not result in a completely flat background as
expected. Instead, patterns remained in the images, probably because
of the rapid and variable atmospheric conditions.  However, these
patterns varied on scales of $\sim$$2"$, that is, on scales larger than
the size of the SAM PSFs. Their impact on the calibration of the SAM
observations was therefore considered not significant. A reduced
individual image frame ($L'$, \textit{BB\_9holes}) is shown in
Fig.\,\ref{fig:samim}.

The so-called cube mode was used to save each single-detector
integration time (DIT) frame. The advantage of this mode is that it
allows selecting the best frames. In our case, frame
  selection was necessary because atmospheric conditions (and the AO
  correction) were affected by highly variable seeing and the passage
  of clouds. About $80\%$ of the frames were de-selected
(using a simple flux criterion on the star IRS\,7).  The final set of
data consists of a group of 13 cubes composed of $20-30$ exposures
with 5\,s and 20\,s DIT each for $L'$ and $NB\_3.74$ filters, respectively. Finally, the selected frames for each of the
individual targets were stored into  separate cubes of $128\times128$ pixels.

The standard observing technique for SAM requires closely interwoven observations between calibrator and target. The calibrator must be a nearby point-like source (within $\sim$$1^{\circ}$-$2^{\circ}$). We obtained observations of the calibrator sources HD\,161718 and HD\,162907. A dither pattern was used in which the source was located at the center of the four quadrants of the detector so that the sky could be extracted by a simple median superposition of the dithered target observations. 

\subsection{Extraction and calibration of the interferometric
  observables from the SAM data}

\begin{figure}
\centering
\includegraphics[width=8 cm]{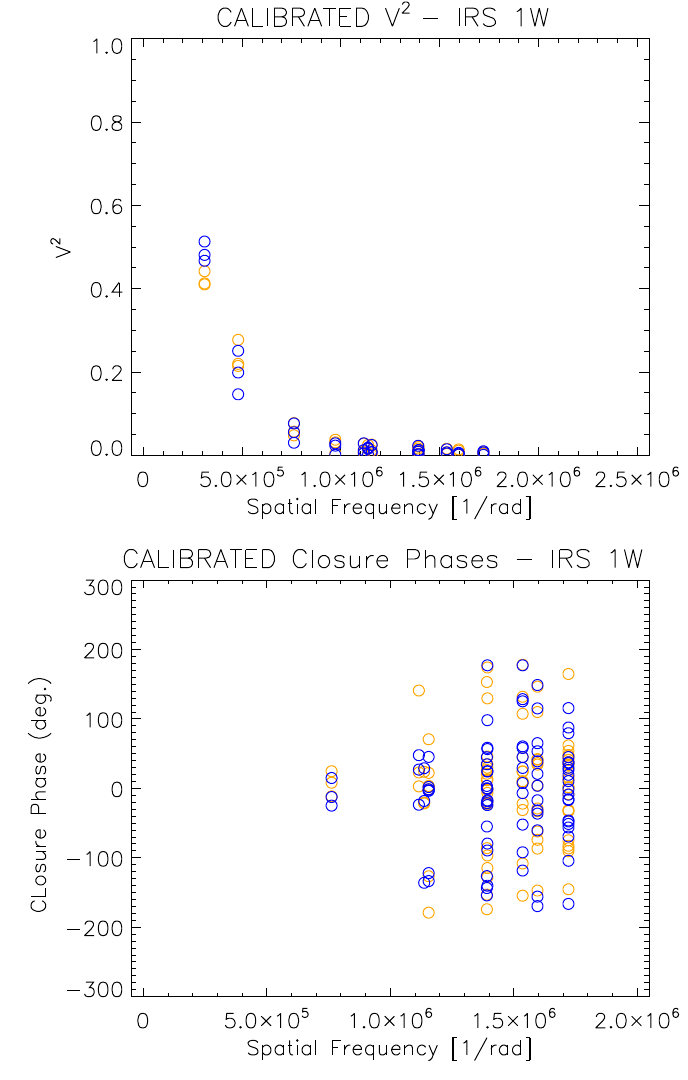} 
\caption{Calibrated squared visibilities and closure phases for the
  SAM observations of IRS\,1W. Orange circles: calibration with the
  synthetic calibrator technique. Blue circles:  calibration with
  IRS\,7 as calibrator. The CPs values are plotted over the spatial frequency
of one of the three baselines used to compute each one of them.}
 \label{fig:calib}
\end{figure}

For the subsequent processing of the SAM data we used an IDL pipeline developed 
at Sydney University (priv. comm., M. Ireland).  This code constructs a matched 
filter for the given instrumental setup (pixel scale, observing wavelength, 
filter, aperture mask) and uses it to measure the squared visibilities, $V^{2}$, 
for all interferometric baselines as well as the bispectra, which
are the ensemble 
of the visibility products for all non-redundant baseline triangles. The 
arguments of these bispectra are the closure phases (CPs). The CPs provide information of the point-symmetry of the source and are free of telescope-dependent phase errors induced by atmosphere and telescope vibrations. Finally, to eliminate the baseline-dependent errors,
the calibrated $V^{2}$ were obtained by the ratio of the raw V$^{2}$ of the 
target and calibrator, while the calibrated CPs were obtained by subtracting 
the calibrator CPs from the CPs of the target.

We note that the $u-v$ coverage ($u-v$ refers to the coordinates in
the interferometric plane, i.e., the coverage of the Fourier transform of the image)
of the targets in our observation was not limited by the properties of
the interferometric masks used, but was increased through Earth
rotation synthesis by combining observations of the targets at
different parallactic angles.

To accurately calibrate the SAM observations, it is generally
recommended to frequently switch between the target and the calibrator
while leaving all AO settings unchanged. But this was not
possible for our observations because of the highly variable
conditions (seeing, clouds) that made changes to the AO setting
necessary, delayed calibrator observations, and led to unstable
correction.  We thus lacked sufficient, adequate calibrator
observations.  

One of the possible solutions to overcome this problem was using
the IRS\,7 star as calibrator because it is contained in all our SAM
frames. This source was already tested and used by
\citet{Tuthill:2006hh} as a calibrator for SAM observations. However, they performed their observations at shorter wavelengths than 
we did. IRS\,7 is known to be surrounded by extended emission 
at MIR wavelengths \citep{Viehmann:2006uq}, which may already show some 
contribution at the long NIR wavelengths used here. In addition, the images of 
IRS\,7 may have been affected by local systematic variations of the 
background and do not properly reflect the variance of PSF and
background for the entire FOV of our observations.  

To overcome these problems we developed a
technique that we called the synthetic calibrator. This method
takes advantage of the many bright sources in a crowded
FOV as is the case for the GC.  The procedure is very similar to
the iterative extraction of an empirical PSF from a crowded field that is implemented in PSF fitting algorithms such as {\it StarFinder}
\citep{Diolaiti:2000fk}. We used {\it StarFinder} to (i) select the
brightest stars in the SAM images (excluding known, obviously extended
sources such as IRS\,1W) to construct a preliminary PSF by median
superposition; (ii) clean the images from secondary sources with the
help of the preliminary PSF and subtract large-scale variations of the
background emission; and (iii) obtain a final PSF from the median
superposition of the cleaned images of the reference stars. The final
PSF, which corresponds to the interferogram of a point source, was
normalized to a total flux of one. We used the sources IRS\,7, IRS\,3,
IRS\,29, and IRS\,6E to construct the synthetic calibrator.

This technique provides a reliable way to minimize the contribution of
all the potential systematic errors related to each one of the
individual sources. Even more, the synthetic calibrator is
robust, under certain constraints, to the use of slightly extended
sources as inputs (like IRS\,3). These restrictions require that the in-going sources
have no strong closure phase information (i.e. the source is not
asymmetric) and that their emission is dominated by the point-like
source. We can observe, at first glance, in Fig.\,\ref{fig:samim}
that the four sources used to build the synthetic calibrator fulfill the requested constraints and have a
significantly crispier PSF than our science targets.

We used the few valid calibrator observations to verify this technique
by comparing the calibrated $V^{2}$ and CP values that were obtained
with the synthetic calibrator and with a real calibrator. No
systematic differences were found. We also compared the resulting
calibrated $V^{2}$ and CP obtained with the synthetic calibrator and
with IRS\,7 as calibrator. Again, the values agreed within their
uncertainties. This also confirms that IRS\,7 is unresolved with the
chosen SAM setup.  Figure\,\ref{fig:calib} shows a comparison of the
calibrated V$^{2}$ and CPs using the synthetic calibrator technique
and IRS\,7 as calibrator for one of our sources.  We decided to use
the synthetic calibrator technique to calibrate the $V^{2}$ and CP of
all our targets. 
Figure\,\ref{fig:calib} illustrates the information that can be obtained 
from our SAM measurements: IRS\,1W is completely resolved-out for
baselines longer than 4 meters, which indicates that the source angular
size is extended and larger than 120 milliarcsec; the closure phases differ 
significantly from zero, which excludes a point-symmetric structure
for the source.

\begin{table}
  \caption{Iterations and reduced $\chi^{2}$  for BSMEM image reconstruction }
\label{tab:bsmem}
\centering
\begin{tabular}{l r r}
\hline\hline
Source & Iterations & $\chi^{2}$ \\
\hline
   IRS 1W & 15 & 1.10 \\
   IRS 5 & 15 & 1.86 \\
   IRS 10W & 12 & 2.31 \\
   %IRS 2L & 9 & 2.52 \\
   IRS 21 & 20 & 1.82 \\
\hline
\end{tabular}
\end{table}

We note that the synthetic calibrator technique can only be used when
several sufficiently bright sources are present within the FOV and
when the PSF is not expected to vary significantly over the FOV. It may
 be ideal for observations of stars in dense clusters, where it can both
improve the calibration of SAM data and boost the efficiency of SAM
observations because it does not require any change of the instrument
setup and avoids overheads caused by calibrator observations.

\subsection{SAM image reconstruction}

Image reconstruction from the interferometric data was performed with
the BSMEM package \citep[see, e.g.,][]{ Buscher_1994, Lawson:2004uq}. This code uses
a maximum-entropy algorithm. Table\,\ref{tab:bsmem} lists the number
of iterations and the reduced $\chi^{2}$ values at the end of the
deconvolution process for all sources. The size of the beam used in
the reconstruction procedure is $\theta$=60\,mas, which corresponds to
the nominal angular resolution reached by this technique for the
filters we used. The pixel scale used for the reconstruction process was 13 mas/px.

\section{Proper motions \label{sec:proper}}

\begin{table*}
\centering
\caption{Measured proper motions in km/s of stars and bow-shock sources.}
\label{Tab:proper} 
\begin{tabular}{lrrrrrr}
\hline
\hline
Name & $v_{R.A.}^{\mathrm{a}}$ & $v_{Dec.}^{\mathrm{a}}$ & $v_{R.A.,S09}^{\mathrm{b}}$ & $v_{Dec.,S09}^{\mathrm{b}}$  & $v_{R.A.,T05}^{\mathrm{c}}$ & $v_{Dec.,T05}^{\mathrm{c}}$ \\ 
\hline
IRS\,1W &  $-164\pm 35$ & $309\pm22$ & $54\pm52$ & $167\pm55$ & $-170\pm30$ & $320\pm60$\\
IRS\,5 &  $54\pm18$ & $26\pm25$ & $57\pm4$ & $32\pm12$ & & \\
IRS\,10W & $-21\pm21$ & $216\pm14$ & $-43\pm5$ & $223\pm7$ & $-160\pm110$ & $10\pm210$\\
%IRS\,2L & $47\pm28$ & $-123\pm16$ & $38\pm10$ & $125\pm9$ & & \\
IRS\,21 & $-165\pm13$ & $57\pm33$ & & & $-300\pm30$ &  $100\pm30$ \\
IRS\,16NW & $225\pm10$ & $26\pm11$ & $195\pm5$ & $40\pm2$ & & \\
IRS\,33N & $110\pm17$ & $-239\pm18$ & $97\pm4$ & $-229\pm2$ & & \\
\hline
\end{tabular}
\begin{list}{}{}
\item[$^{\mathrm{a}}$] Proper motion measured in this work.
\item[$^{\mathrm{b}}$] Proper motion measured by \citet{Schodel:2009zr}.
\item[$^{\mathrm{b}}$] Proper motion measured by \citet{Tanner:2005fk}.
\end{list}
 \end{table*}\label{tab:proper_motion}

 To measure the proper motions of the stellar sources at the center of
 the bow shocks we used VLT/NACO $K_{s}$-band
 AO imaging data obtained between 2002 and 2011 (see
 Table\,\ref{Tab:ImObs}). \citet{Schodel:2009zr} obtained
 proper motion measurements of our sources using PSF fitting to
 measure their stellar positions. This method is not suited for extended sources like the
 bow shocks that we analyze here, however. Instead, we measured the stellar
 positions by centroiding the emission maxima of the central stars in circular apertures of with a radius  of 4, 5, and 6 pixels. The positions and their uncertainties were taken as the mean
 of the three measurements and its corresponding uncertainty. Note
 that measuring stellar positions by aperture centroiding is
 significantly less susceptible to systematic errors caused by
 (moderate) saturation or nonlinearity than is PSF fitting.  Moreover, since the
 bow shocks are extended targets, PSF fitting to their central stars
 is very difficult and prone to systematic errors.

 Astrometric alignment was achieved with the help of the known
 positions and proper motions of a sample of about 200 stars contained
 in all images. The astrometric reference stars were selected from the
 list published by \citet{Schodel:2009zr}. The stars had to be brighter than $K_s=12$, with no
 companion within $0.3"$ and the uncertainty of each component
 of the proper motions of the reference stars had to be smaller than
 $<30$\,km\,s$^{-1}$. These criteria ensure accurate aperture measurements of their positions.

We transformed the stellar positions into the astrometric reference
frame via a second-order polynomial fit \citep[see Appendix A
in][]{Schodel:2009zr}. A Monte Carlo
approach was used to estimate the uncertainty of this alignment
procedure, with 1000 trials. For each trial the positions of
the reference stars were varied randomly, with the offsets drawn from
Gaussian distributions with standard deviations corresponding to the
positional uncertainty of each star at each epoch.

After obtaining the stellar positions for each epoch, we fitted the
data points by linear motions. The reduced $\chi^{2}$ was generally
significantly larger than 1. There were some sources
of uncertainty that were not specifically taken into account,
such as errors introduced by camera distortion into the mosaics,
the shape of the bow shocks, pixel saturation, or faint secondary
stars within the aperture laid over the bow shocks. Hence, the
uncertainties of the stellar positions may have been
under-estimated. This is no serious problem, however, because we can
be certain that linear motions are accurate models to describe the
proper motions. Therefore, the uncertainties of the linear fits were
all rescaled to a reduced $\chi^{2}=1$. This procedure avoids
underestimating the uncertainties (see Sch\"odel et al., 2009, for a
more elaborate discussion). The measured angular velocities were
converted into km\,$s^{-1}$ by assuming a Galactic center distance of
8\,kpc \citep[e.g.,][]{Schodel:2010fk}. Plots of position vs.\ time
along with the linear fits for the proper motions are shown in
Appendix\,\ref{sec:propplots}.

Table\,\ref{Tab:proper} lists the proper motions of the bow-shock 
sources and of the stars IRS\,16NW and IRS\,33N as determined here and 
by \citet{Schodel:2009zr}. IRS\,16NW and IRS\,33N are two point-like
sources that we used as a reference point to compare our method and \citet{Schodel:2009zr}. The measurements of IRS\,16NW and IRS\,33N and of the bow-shock sources all agree 
within their $1\,\sigma$ uncertainties, with the exception of 
IRS\,1W. We believe that the discrepancy on the latter source 
is probably due to the PSF fitting technique used 
by \citet{Schodel:2009zr}, which is pooröly suited to strongly 
resolved/extended sources like IRS\,1W.  Moreover, IRS\,1W is one of 
the brightest sources in the field and is affected by saturation 
in many images. Aperture fitting is suited to both extended and 
(moderately) saturated sources. We therefore believe that the 
proper motions for the bow-shocks' stellar sources that we have derived 
here may be more accurate than those published 
by \citet{Schodel:2009zr}. \citet{Tanner:2005fk} measured the 
proper motions of IRS\,21, IRS\,1W, and IRS\,10W, which are 
also listed in Table\,\ref{Tab:proper}. Their measurements 
generally agree with ours within the quoted $1\,\sigma$ 
uncertainties, with the exception of the velocity along the 
right ascension of IRS\,21. Note that the proper motions 
of \citet{Tanner:2005fk} suffer from very large uncertainties, 
possibly because of the small FOV that they used for 
determining the proper motions and because they used images 
reconstructed from speckle observations that have a significantly 
lower signal-to-noise ratio than we used 
(AO imaging).

\section{Modeling the bow shocks \label{sec:BSmodel}}

\subsection{Bow-shock images \label{sec:bs_images}}

\begin{figure}
\centering
\includegraphics[width=\columnwidth]{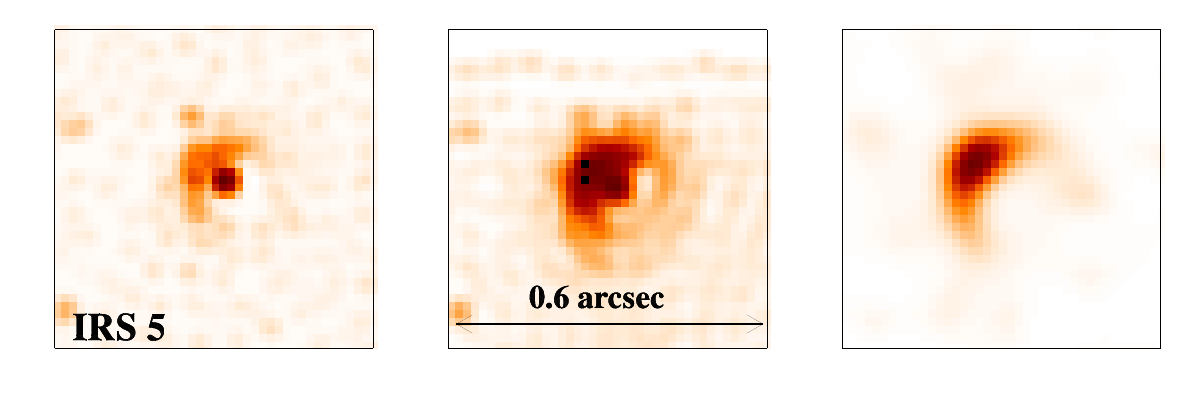} 
\includegraphics[width=\columnwidth]{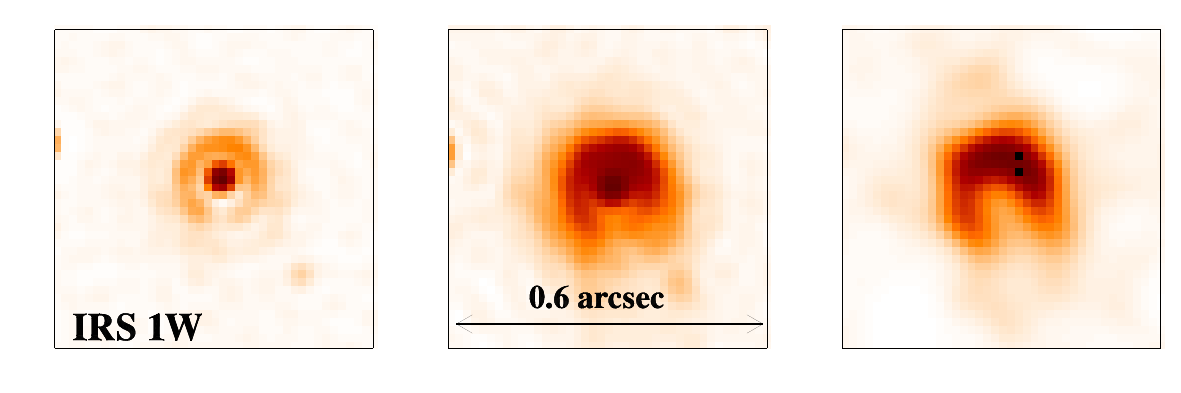} 
\includegraphics[width=\columnwidth]{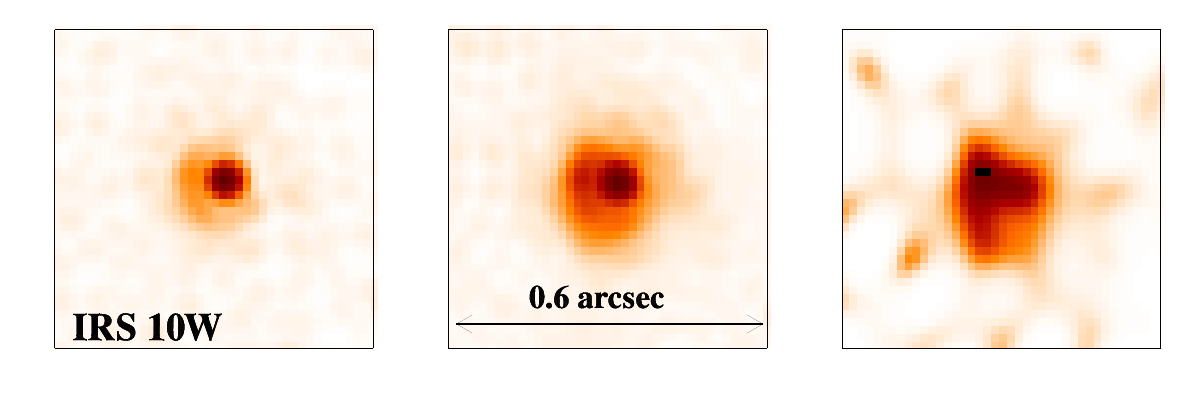} 
\includegraphics[width=\columnwidth]{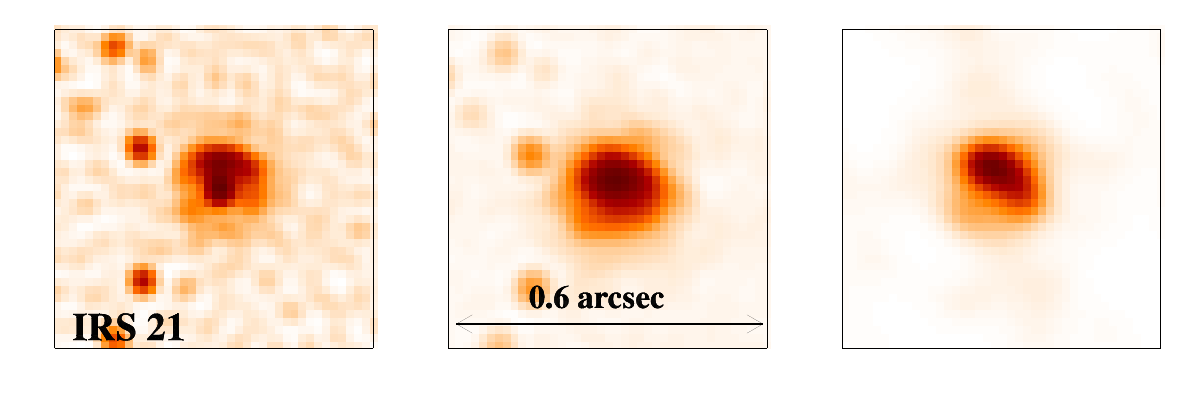} 
\caption{Images of the bow-shock sources. Each line corresponds to a different
  source, while the three columns show the Lucy-Richardson and
  beam-restored images in the $H$- and $K{s}$-bands, and the image
  reconstructed from the  $L'$-band SAM observations.}
 \label{fig:BS-images}
\end{figure}

In Fig.\,\ref{fig:BS-images} we show a compilation of our images of
the bow-shock sources IRS\,1W, IRS\,5, IRS\,10W, and
IRS\,21. The bow shock is clearly resolved in almost all sources in at least
one wavelength. IRS\,21 and
IRS\,10W are more clearly resolved than in the work of
\citet{Tanner:2005fk}, which can be attributed to technological
progress. While \citet{Tanner:2005fk} used low Strehl shift-and-add
reconstructed speckle data, here we use high Strehl and high
signal-to-noise AO images and images reconstructed from AO assisted
sparse aperture masking, a technique that provides a precise
calibration of the PSFs. Note that \citet{Buchholz:2011fk} and \citet{Buchholz_2013}
presented a resolved image of the bow shock around IRS\,21. Hence,
this source is not seen face-on as hypothesized by
\citet{Tanner:2005fk} \citep[see also][]{Tanner:2002kx}.

The extended emission becomes stronger toward longer
wavelengths, a consequence of the temperature of the dust in the
bow shocks. \citet{Tanner:2002kx} derived a dust temperature of
$\sim$$1000\,$K for IRS\,21.  \citet{Viehmann:2006uq} also showed that
the temperatures of the bow-shock sources peak at around $4$-$10\,\mu$m,
consistent with a dust temperature of a few hundred degrees. On the
other hand, the central star is only detected at the shorter 
wavelengths (see Fig.\,\ref{fig:BS-images}).

\subsection{Bow-shock shape\label{subsec:shape}}

The shape of the bow shocks results from the interaction between the
stellar wind and the surrounding ISM, taking into account the relative
motion of the star with respect to the ISM.  Assuming a steady-state
model, the shape of the bow shock results directly from the stand-off
distance, $R_{0}$, where the ram pressure from the relative motion
through the ISM is balanced by the pressure from the stellar wind,
 \begin{equation}
   R_{0}=\sqrt{\frac{\dot{m}_{*}\nu_{w}}{4\pi\rho_{\rm ISM}\nu_{*}^{2}}}, 
\label{eq:sd} 
\end{equation}
where $\dot{m}_{*}$ is the mass-loss rate of the star, $\nu_{w}$ is
the velocity of the stellar wind, $\rho_{\rm ISM}$ is the density of the
surrounding ISM, and $ \nu_{*}$ is the relative velocity between the
star and the ISM.

\begin{figure}
\centering
\includegraphics[width=\columnwidth]{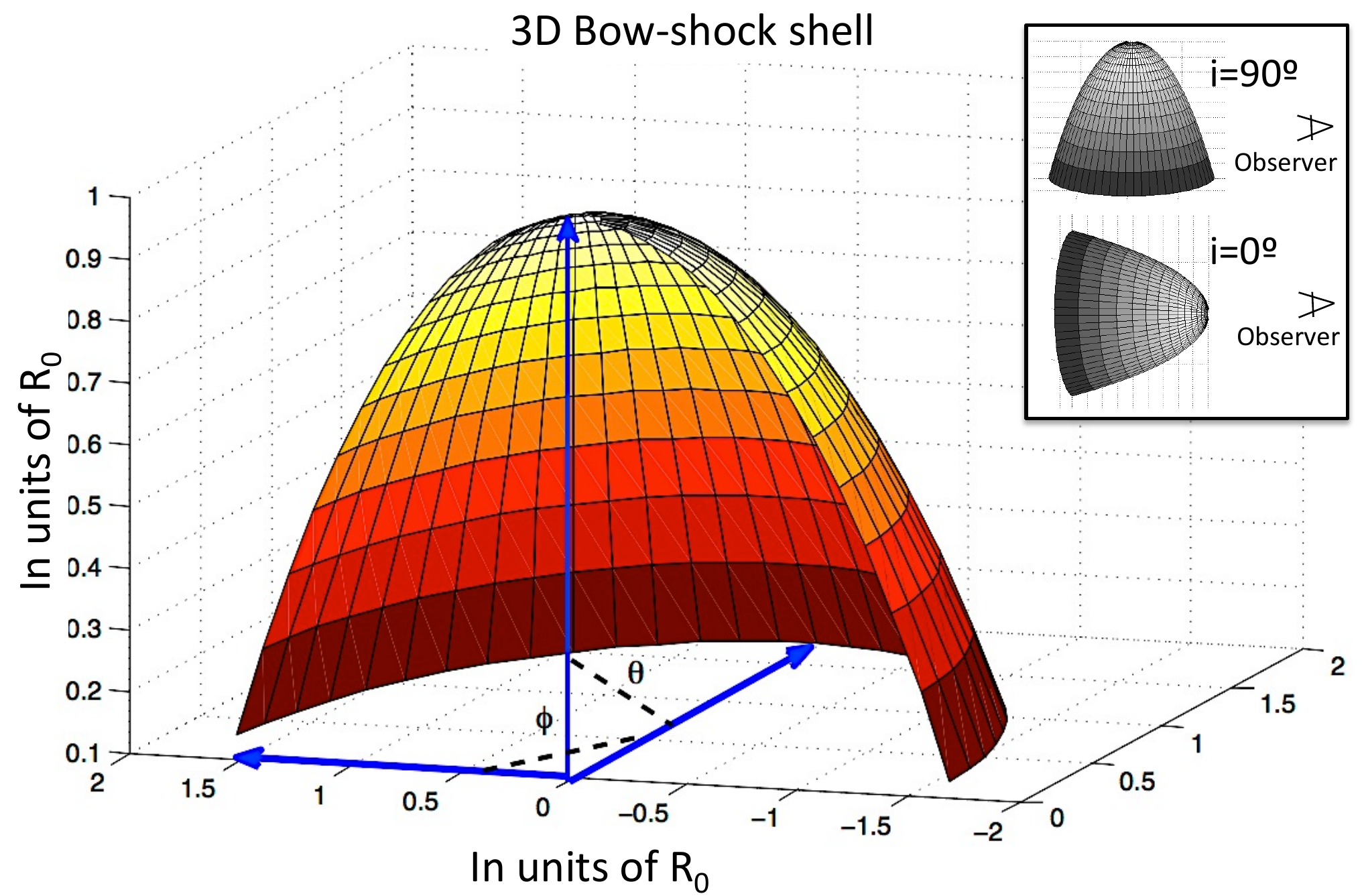} 
\caption{3D Diagram of the bow-shock shell created
by our model. The diagram is sliced and the interior of the bow
shock
is visible. The two polar angles $\phi$ and $\theta$ are
shown (see Eqs.\,\ref{eq:bs_2D}\, and \ref{eq:bs_3D}). The blue axis is centered on the position of the star and the
scale is in multiples of $R_{0}$. The inset in the upper-right part of
the diagram shows the change in bow shock inclination angle as
seen from the observer.}
 \label{fig:BS_shell}
\end{figure}

To obtain a steady-state solution for the bow-shock
shell, \citet{Wilkin:1996kx} assumesed (i) that the bow-shock shape is
determined by the momentum of the shocking winds, (ii) that the
cooling process is instantaneous, (iii) that the shell is a thin layer
of negligible width, and (iv) that the interacting winds are
homogeneous and do not change with time. The 2D polar canonical form of a bow shock, under the previous conditions, is then defined by \begin{equation}
      R(\theta)=R_{0}\csc{\theta}\sqrt{3(1-{\theta}{\cot{\theta}})}, 
\label{eq:bs_2D} 
\end{equation}
 where $\theta$ is defined as the angle from the axis of symmetry to
 any point on the 2D-shell's surface. To extend the previous relation
from 2D to 3D, we let the locus of R($\theta$) rotate by angle $\phi$ as shown in Fig.\,\ref{fig:BS_shell}, hence
\begin{equation}
      R_{3D}(\theta, \phi)=rot \{ R(\theta), \phi \}, \, \, \, \, \,
      \,  0^{\circ} \le \phi \le 180^{\circ}
\label{eq:bs_3D}
 .\end{equation}

The cells of the bow shock are defined by sampling $R_{3D}(\theta,
\phi)$ with increments in $\theta$ and $\phi$ of 0.5$^{\circ}$ (see
Fig.\, \ref{fig:BS_shell} ).

\subsection{Bow-shock emission}

\subsubsection{Polarization and emission mechanisms}

\citet{Buchholz_2013} described that the bow-shock sources,
particularly IRS 21 and IRS\,1W, exhibit strong polarization
levels at NIR wavelengths. The polarization increases from shorter to longer
wavelengths, highlighting the importance of the extended dust
component. Polarization of the extended emission and of the embedded
bow-shock sources in the Northern Arm (NA) reaches its maximum in the
MIR \citep[see][]{Glasse_2003, Buchholz:2011fk,Buchholz_2013}. \citet{Buchholz:2011fk} found that the polarization of the
bow-shock sources IRS21 and IRS\,1W agrees with the orientation of the
magnetic field in the NA. These authors also suggested that the ordered
magnetic field is responsible for the dust grain alignment in the
bow-shock shells. The polarization measurements from NIR to
MIR thus support the assumption that the emission in the bow
shocks
is produced by the dust IR thermal emission, which is reprocessed UV light
from the central star.
  
\subsubsection{Bow-shock emission model}

Although the previously mentioned findings suggest that the bow-shock
emission is mainly produced by thermal emission, we also
explored the role of scattered light in the bow-shock emission,
particularly at shorter wavelengths. The procedure used to simulate
the emission maps was similar to the one developed by
\citet{Muzic_2010}. The dust thermal emission, $L_{th}$, was assumed to have a
blackbody source function attenuated by the corresponding optical
depth, that is, $L_{th} \propto
B(T_{d})(1-e^{-\tau_{abs}})$. The dust temperature distribution in the bow shock
($T_{d}$) is assumed to be created by heating the dust by the stellar
photospheric UV photons from the central source. Therefore, under this condition, $T_{d}$
is computed via the \citet{VanBuren_1988} solution,

 %The effects of the
%shock heating from the collision fo the ISM with the stellar wind are
%beyond the scope of our model and thus, they are not
%considered. 

 \begin{equation}
 T_{d} = 27 a_{\mu m}^{-1/6} L_{*,38}^{1/6} r_{pc}^{-1/3} K
  \,,
\label{eq4}
  \end{equation}

where $a_{\mu m}$ is
the grain size in microns, $L_{*,38}$ is the star UV luminosity in
units of $10^{38}erg/s$ and $r_{pc}$ is the distance to the dust from the
star in parsecs. After calculating the contribution from each cell, the total $L_{th}$ was normalized to one.

The scattering emission was  computed using the normalized \citet{Henyey_1941}
source function modified by the optical depth, $L_{sca} \propto 
r_{\rm cm}^{-2}P(\theta_{sca})e^{-\tau_{sca}}$, where $r_{\rm cm}$ is the distance to
the dust from the star in centimeters. The scattered light is described by

  \begin{equation}
  P(\theta_{sca})=\frac{(1-g^2)}{(1+g^2-2gcos(\theta_{sca}))^{3/2}}
  \,,
  \end{equation}

where $g$ is a dimensionless term that depends on the scattering
  angle $\theta_{sca}$, with $g=0$ when the scattering is fully
  isotropic and $g<1$ when the scattered light preserves the angle of
  incidence. After calculating the contribution from each cell, the
  total $L_{sca}$ was normalized to one. The previous treatment of the
  scattering emission does not take
  into account the polarization of light, only the deflection angle of
  the incident radiation in the dust. A more detailed model of the scattering emission is beyond the
  scope of this work.

The total bow-shock emission, $L_{tot}$, at every cell in the model was
computed adding the previously normalized
thermal and scattering emissions, $L_{tot}=\epsilon _{sca}
L_{sca}+\epsilon _{th} L_{th}$. The coefficients $\epsilon_{th}$ and
$\epsilon_{sca}$ are the relative efficiencies ($\epsilon_{th}+\epsilon_{sca}=1$) of the thermal and scattering contribution. The optical depths, $\tau_{abs}(\lambda)$ and
$\tau_{sca}(\lambda)$, at every cell of the 3D bow-shock shell are calculated as

\begin{equation}
\tau_{abs}(\lambda)=\int_{a-}^{a+}n_d(a)C_{abs}(a,\lambda)da
\,,
\label{eq6}
\end{equation}

\begin{equation}
\tau_{sca}(\lambda)=\int_{a-}^{a+}n_d(a)C_{sca}(a,\lambda)da
\,,
\label{eq7}
\end{equation}

where $C_{abs}(a, \lambda)$ and $C_{sca}(a, \lambda)$
are the dust thermal and scattering extinction coefficients. They are
calculated as $C_{abs}(a, \lambda)=\pi a^{2}(Q_{abs})$ and
$C_{sca}(a, \lambda)=\pi a^{2}(Q_{sca})$. The values of the thermal and
scattering efficiencies,  
$Q_{abs}$ and $Q_{sca}$,  were obtained
form the Draine database on the optical properties of interstellar
dust grains\footnote{http://www.astro.princeton.edu/~draine/dust/dust.diel.html} for a standard mixture
of 50º\% graphite and 50\% silicate \citep[see][]{Mathis:1990uf, Draine_1984}. The grain
size distribution used, $n_{d}(a)$, follows an MRN \citep{Mathis_1977} power-law with the grain
size $a$ ranging from $a-$=0.005$\mu$m to $a+$=0.25$\mu$m.

\subsection{Bow-shock fitting  \label{bs_fitting}}

We used the 3D steady-state model and the dust emission
processes described in the two preceding sections to create models of
the bow-shock shells and fit them to the images. The best-fitting
models were found by a $\chi^{2}$-minimization as implemented in the
IDL {\it MPFIT} package by C. Markwardt \citep{Markwardt_2009}. Masks were applied to the
bow-shock images to suppress features that are not related to the
actual bow shocks. Each masked image was normalized to unity inside the
applied mask. The same normalization was applied to the projected and masked models. A central point-source was included in the model
(with free parameters for its position and flux). 
This point-like source concentrates most of the NIR-flux and dominates the
model fitting at the shortest wavelengths.
We note that the
pixel uncertainties are hard (or impossible) to determine in
Lucy-Richardson deconvolved images and in the BSMEM reconstructed
images. In addition, there are several sources of systematic uncertainties,
such as faint point sources or ISM features that may be confused with the
bow shocks. Therefore, both uniform and flux-weighting schemes were tested to determine the model with the lowest uncertainties that is least
affected by the systematics previously discussed. Our tests
show that a flux-weighting scheme produced better fits in $H$ and
$K_s$ while a uniform weighting works better in $L'$. The final
uncertainties of the best-fit parameters are obtained from their
standard deviation of the fits to the
statistically independent images from the three different filters. We
also expect this procedure to reliably take care of possible
contamination from stars and extended emission because the former are
expected to be faint at $L'$ and the latter at $H$.

\begin{figure}
\centering
\includegraphics[width=\columnwidth]{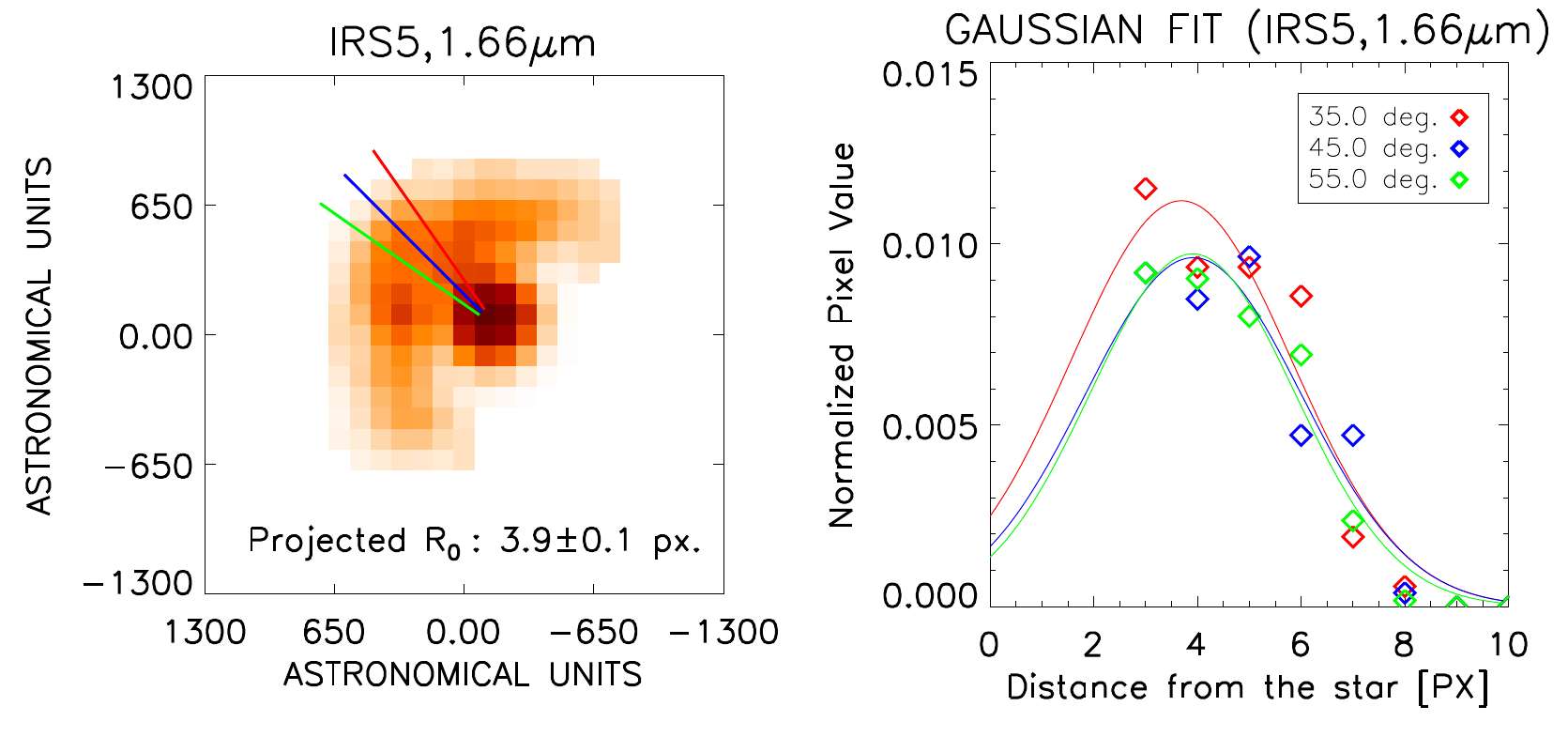} 
\caption{\textbf{Left:} $H-$band image of IRS5 with the three radial cuts used
  to estimate $R_{0}^{'}$
  marked in colors. \textbf{Right:} Gaussian fitted to the emission of the
  bow shock in the direction of the three radial cuts. The
  position angle (E of N), used for each cut, is identified in different colors.}
 \label{Fig:radial}
\end{figure}

The stand-off distance $R_0$ (reported in Table\,
\ref{model:irs5}), which is
strongly degenerated with the inclination of the bow shock, was linked in our fitting code to  the projected stand-off distance in
the plane of the sky, $R_{0}^{'}=R_{0}/cos(i-90^{\circ})$, where $i$ is the
bow-shock inclination angle between the LOS and the bow-shock
apex. When the bow-shock apex is pointing to the
observer $i=0^{\circ}$ and when the bow-shock shell is projected in the plane
of the sky $i=90^{\circ}$ (see inset in Fig.\,\ref{fig:BS_shell}). $R_{0}^{'}$ was obtained from the mean of a Gaussian fit to the
emission obtained by three radial
cuts along the bow-shock structure. The central cut was made along
the symmetry axis of the bow shock, the two additional cuts were at $-10^{\circ}$ and $+10^{\circ}$ from the position
of the central cut. Figure\, \ref{Fig:radial} shows as an example the radial cuts, the
fitted Gaussian, and the $R_{0}^{'}$ mean value obtained for IRS5 in
the $H-$filter. A similar procedure was performed for the other
sources. The rest of the parameters
described in Table\, \ref{model:irs5} were fitted directly to the image described at the beginning of this section.   

\begin{figure}
\centering
\includegraphics[width=\columnwidth]{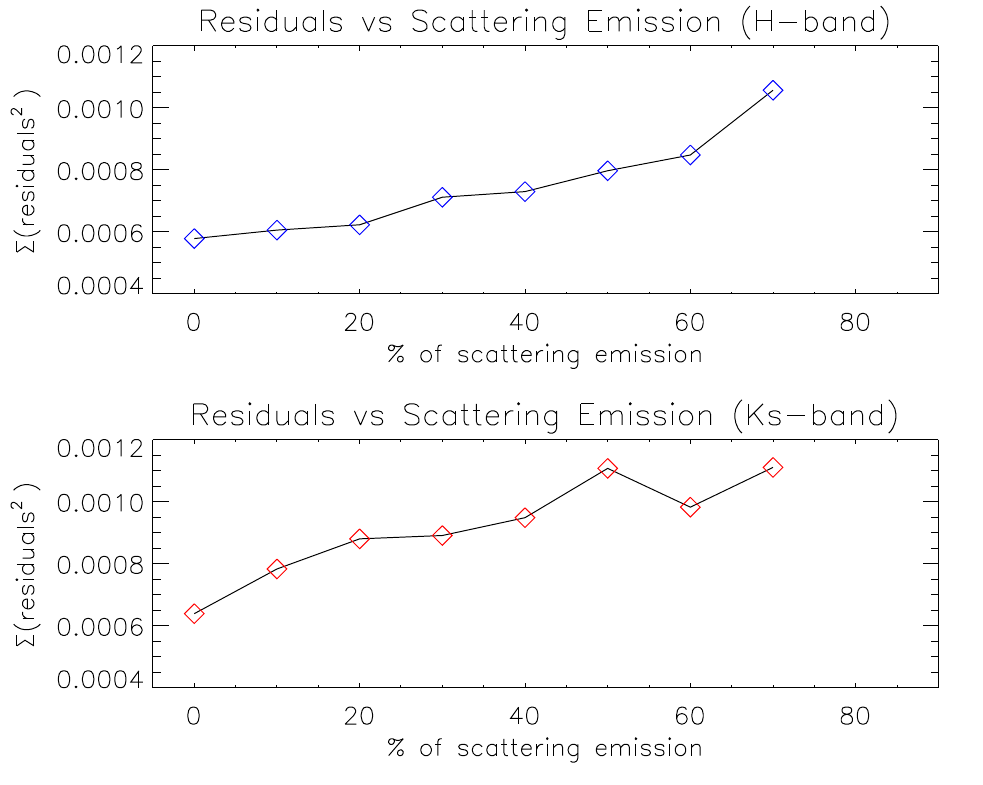} 
\caption{Sum of the squared residuals of the best-fit models with increasing relative contribution of scattering
  emission. The upper panel corresponds to the residuals in the $H   $ filter
and the lower one to the residuals in the $K_s $ filter for
isotropic scattering.}
 \label{fig:residuals}
\end{figure}

\subsection{ Contribution from dust scattering to the bow-shock emission \label{scat_vs_therm}}

To quantitatively approximate their relative importance, we investigated the effect of different
relative thermal and scattering efficiencies on our models. Since IRS
5 appears to resemble the canonical steady-state
bow-shock model most, we used the images of this source in $H$ and $K_s$ to perform this
test. We fit different sets of models with fixed efficiencies for thermal
and scattering emission at different wavelengths and considered cases
of isotropic and anisotropic scattering. For each wavelength, model
fitting was performed with increments of 10\% in the scattering efficiency
from 0 to 70\% of the total emission
observed. We tested the cases for g=0 in the scattering source
function (isotropic scattering) and for g=0.5 (partially anisotropic
scattering). 

The best-fit bow-shock model was found in each trial with the
procedure described in Sect.\,\ref{bs_fitting}. Residual images were
obtained by subtracting the models from the original image. Figure\,\ref{fig:residuals}
displays the sum of the squared pixel values of the residual image for
each trial in $H$ and $K_s$ for the isotropic case. In both
bands, we found that the
residuals increase with the scattering contribution, the model
with 0\% scattering had the lowest residuals. Similar
results were found for the anisotropic scattering contribution (g=0.5). As
indicated in Sect. \ref{sec:bs_images}, we expect the main
contribution to the bow-shock emission to be thermal emission from the
dust grains. Our tests here are consistent with this
picture. Therefore, all the models obtained for this work were fit only for the thermal
emission of the dust.

\subsection{Properties of the individual sources \label{sec:results}}

\begin{figure}
\centering
\includegraphics[width=\columnwidth]{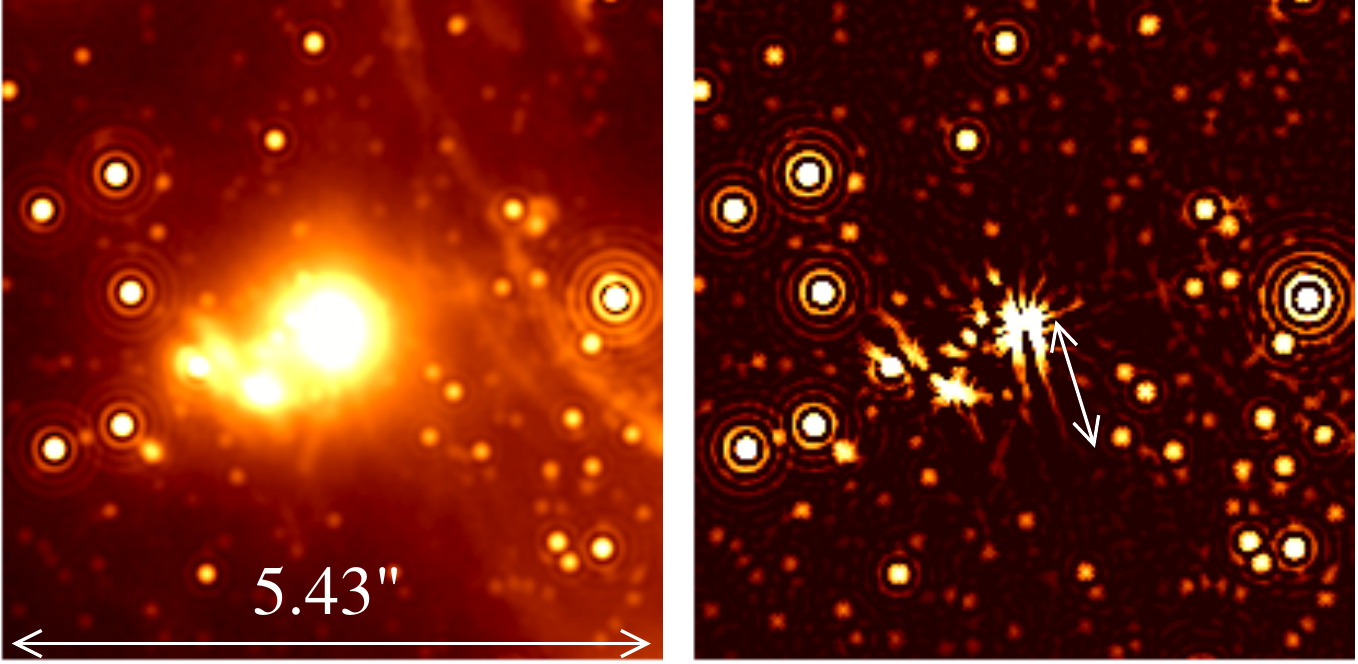} 
\caption{Left: $L'$-band image of the surroundings of IRS\,1W. Right:
  highpass-filtered version of the image. The white arrow in the right
  panel shows the length of the bow-shock tail. It has a length of
  $\sim $ 0.8", or $6400$\,AU. The radial lines emanating from the apex of the bow shock of IRS\,1W are artifacts of the high-pass filtering.}
 \label{fig:irs1w}
\end{figure}

Here, we discuss the individual bow-shock sources in detail. Their
morphologies can be examined in
Fig.\,\ref{fig:BS-images}, while Fig.\,\ref{fig:models} shows the best-fit bow-shock models overplotted to the images. 

{\bf IRS\,5:} This bright bow-shock displays similar morphologies in the
$H$, $K_s,$ and $L'$ filters. The central star is clearly
identified in $H$ and $K_s$. The projected bow-shock shape is
axially symmetrical in the plane of the sky with the vector from
the central star to the stagnation point. IRS\,5 is the source in our sample that
can best be described with the isotropic wind steady-state model. Table\, \ref{model:irs5}
describes the best-fit parameters for the different wavelengths. Here,
and for the other bow-shock sources, we conservatively estimate
the uncertainties of the best-fit parameters from their standard
deviation that results from the fits to the images at the different
wavelengths (i.e., we did not use the uncertainty of the mean). The
small uncertainties support the model assumption of a bow shock in
the steady-state phase. Our results suggest that the bow shock of IRS\,5 is inclined at
$\sim$130$^{\circ}$. 

From the interaction of IRS\,5 with the NA and using the NA model of
\citet{Paumard:2004kx}, we estimate a position of z=3.81''$\pm$2.66'' along
the line-of-sight (LOS), where z=0 corresponds to the plane of the
sky. Assuming that IRS\,5 is gravitationally bound to Sgr\,A*, we
calculate a $3\,\sigma$ upper limit of $\pm$336 km s$^{-1}$ for its LOS velocity.

\begin{table}[htp]
\caption{Best-fit parameters of our bow-shock models for IRS\,5
  (spherically symmetric stellar wind)\label{model:irs5}} % title of Table
\centering  % used for centering table
\begin{tabular}{l c c c c } % centered columns (5 columns)
\hline
Parameter & $H$ &$K_s$ & $L'$ &$\sigma_{rms}$\\ 
\hline
\hline                  % inserts single horizontal line
$R_{0}$ [AU] & 445.0 & 517.0 & 575.0 & $\pm53.0^{\mathrm{a}}$\\ % inserting body of the table
P.A. [deg]$^{\mathrm{b}}$ & 50.0 & 46.9 & 47.0 & $\pm1.6^{\mathrm{a}}$\\ 
Inclination [deg]$^{\mathrm{c}}$ & 116.0 & 139.9 & 144.6 & $\pm12.5^{\mathrm{a}}$\\
FWHM [px]$^{\mathrm{d}}$ & 2.0 & 2.23 & 4.6 & - \\
Bow-shock flux [\%]$^{\mathrm{e}}$ & 60.0 & 90.0 & 100.0 & - \\ 
\hline
\end{tabular}
\begin{list}{}{}
\footnotesize {\item[$^{\mathrm{a}}$] The standard deviation of the
  measurements obtained in the three filters.}
\footnotesize{\item[$^{\mathrm{b}}$] Position angle increasing E of N.}
\footnotesize{\item[$^{\mathrm{c}}$] Angle between the the bow-shock
  symmetry axis and the observer's line of sight. $0^{\circ}$ - the
  bow-shock apex is toward the observer, $90^{\circ}$ - the bow-shock
  is projected on the plane of the sky.}
\footnotesize{\item[$^{\mathrm{d}}$] Full width at half maximum of the
  beam at different wavelengths in pixels. The FWHM reported for the
  $L'-$band is fixed and corresponds to the synthesized beam of the
  SAM observations.}
\footnotesize{\item[$^{\mathrm{e}}$] Percentage of the total
  emission that corresponds to the extended emission. The rest of the
  emission corresponds to the central source.}  
\end{list}
\end{table}

\begin{figure}
\centering
\includegraphics[width=7.0cm]{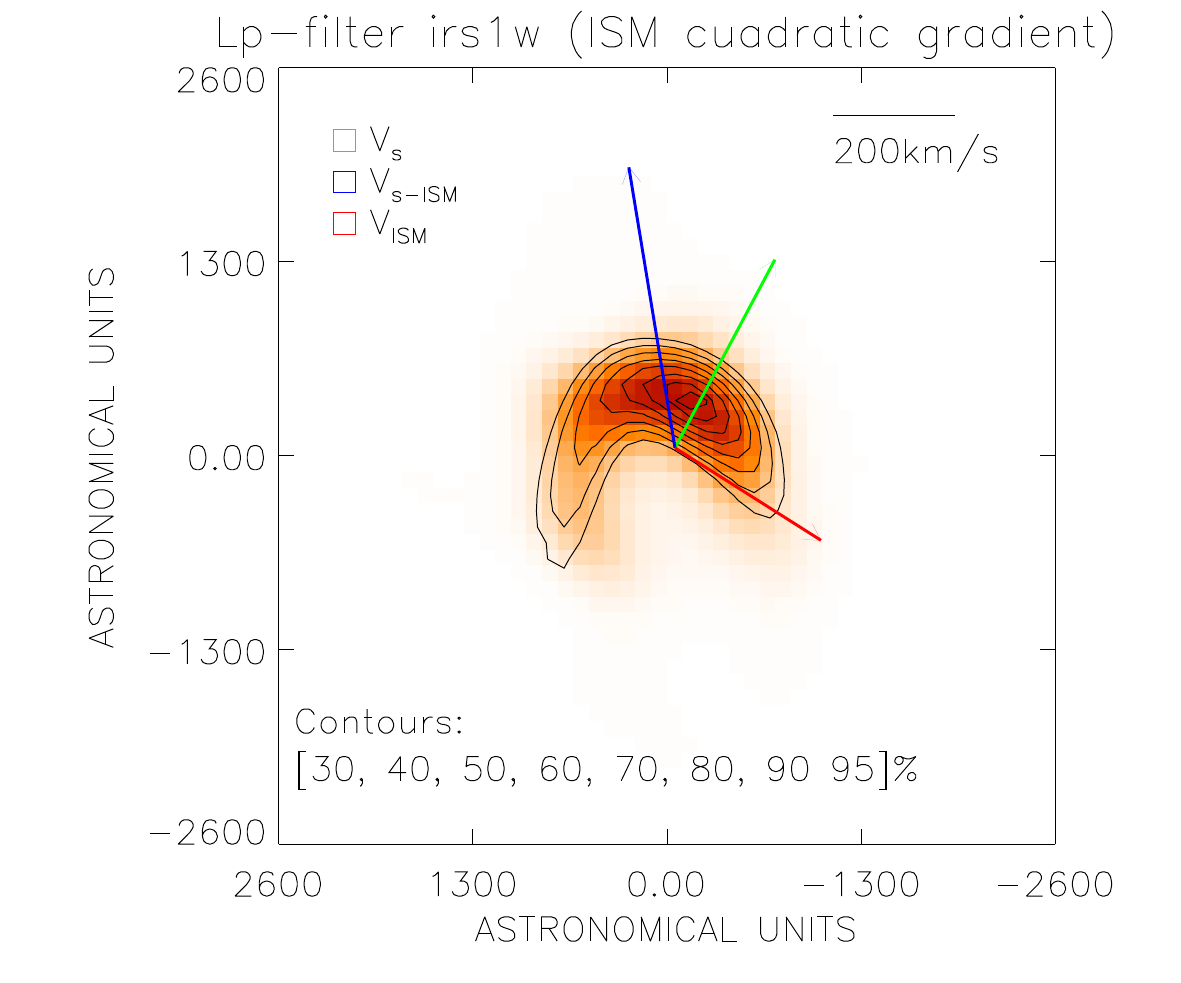} 
\caption{Model of a bow shock in an ISM with a quadratic density gradient. The $L'$ image of
  IRS\,1W is represented in color and the best-fit gradient model is overplotted as contours. The colored vectors represent
    the velocity of the star ($V_s$; green arrow), the relative motion of the star
    through the ISM ($V_{s-ISM}$; blue arrow) and the dust motion of the ISM
    obtained with our bow-shock model ($V_{ISM}$; red arrow).}
 \label{fig:irs1w_grad}
\end{figure}

{\bf IRS\,1W:} as Fig.\,\ref{fig:BS-images} shows, the target is
resolved at the three wavelengths. It has a clear
bow-shock structure in $K_s$ and $L'$, but also deviates from
the analytical bow-shock model that we used here: first, a rounded shell appears around the central star that is open on its
southern edge and is clearly visible at $H$ and $K_s$. It is
not observed in
$L'$ , however. Instead, the tails of the bow shock are better
distinguished. The origin of this more rounded shell in the $H$ and
$K_s$ images is unclear.  Second, the shape near the apex is slightly
asymmetrical in $L' $, with the maximum of the emission shifted to the
west of the apex. Possible causes are confusion with a star
or  some feature in the mini-spiral, or an asymmetrical outflow from
IRS\,1W. To investigate this point further, the left panel in
Fig.\,\ref{fig:irs1w} shows a close-up of the surroundings of IRS\,1W
from a NACO $L'$-band image. The right panel shows a highpass-filtered
version of the same image. IRS\,1W does not appear to be contaminated
by a typical mini-spiral feature such as a thin filament, nor by any
bright star (see also Fig.\,\ref{fig:BS-images}). We therefore
do not find
any obvious or easily testable explanation for the observed asymmetry,
but note here that the mini-spiral is generally a complex
region with many knots and features in the ISM that may be confused in projection and/or interact with the bow shock. Third, IRS\,1W displays a long, tightly closed tail.  The tail of IRS\,1W is clearly visible in the highpass-filtered image and has a length of approximately $0.8"$, or $6400$\,AU. 

IRS\,1W shows a remarkable interaction with the ISM of the NA. 
The proper motion of IRS\,1W is
$350\pm25$\,km\,s$^{-1}$ at an angle $332\pm5^{\circ}$ E of
N. Assuming a mass of $4.2\pm0.3\times10^{6}\,M_{\odot}$  for SgrA*
\citep[][]{Gillessen:2009qe,Yelda_2011}, the escape velocity at the
projected distance, $R=5.30\pm0.01"$, of IRS\,1W is
$419\pm15$\,km\,s$^{-1}$. The LOS distance of IRS\,1W can be estimated
from its interaction with the Northern Arm of the mini-spiral. Using
the model of \citet{Paumard:2004kx}, we estimate a LOS distance of
$z=7.37\pm1.5"$ for IRS\,1W and obtain an escape velocity of
$320\pm29$\,km\,s$^{-1}$  at its 3D position. Although
this is lower than its proper motion, both values overlap well within
their $1\,\sigma$ uncertainties. Using the $3\,\sigma$ upper limit in
the escape velocity and the $3\,\sigma$ lower limit on the proper
motion, we can set a firm upper limit of $V_{z}=300\,$km\,s$^{-1}$ on
the LOS velocity of IRS\,1W (assuming that it is bound to Sgr\,A*, of
course). This corresponds to a maximum inclination of its velocity
vector of $137^{\circ}$, corresponding to $47^{\circ}$ out of the
plane of the sky, moving away from the observer. Here, the ambiguity
between motion toward or away from the observer is broken by our
best-fit bow-shock model. From the model of \citet{Paumard:2004kx} we
can calculate the properties of the gas flow. At the position of
IRS\,1W, the gas in the NA has a proper motion of
$226\pm70$\,km\,s$^{-1}$ at an angle $207\pm13^{\circ}$ E of N. The
orientation of the tail-like feature agrees well with the relative
proper motion between star and gas. The angle of the gas flow out of
the plane of the sky is a mere $7^{\circ}$, with the gas moving toward
the observer, which means that the NA flow is practically within the plane of the
sky at the position of IRS\,1W.  We note that with a relative proper
motion between star and gas on the order of  $550$\,km\,s$^{-1}$  the
channel-like feature that is marked by the tail of IRS\,1W may trace
the motion of IRS\,1W  over the past $\sim$50\,years. With such short
timescales at work, IRS\,1W may be an ideal probe for the conditions
of the ISM in the NA, for instance, for the homogeneity of the gas density.

Table\, \ref{tab:irs1w_1} shows our best-fit bow-shock model for IRS\,
1W using our steady-state model. The second row of images in Fig.\,
\ref{fig:models} corresponds to the steady-state isotropic wind-fitted models. As can be clearly seen, our model cannot fit the
tail of IRS\,1W well (specially in $L'$) because it is much narrower than what can be produced
with this model. Possible explanations of this problem include (i) changes
in the density of the ISM at the shocking points, (ii) variations
in the mass-loss rate and wind velocity, (iii) anisotropic winds, or
(iv) that because of the short interaction times the bow shock in IRS\,1W
has not yet reached the steady-state bow-shock solution
\citep[see][]{Mohamed_2012}. The large scatter in the best-fit
position angles in Table\,\ref{tab:irs1w_1} also suggests that the isotropic wind model is not
suitable to reproduce the observed morphology of IRS\,1W.

To examine whether there are other models that provide a better fit to the IRS\,1W morphology, we decided to
perform a model fitting using the narrow
solution of a steady-state bow-shock shell described in
\citet{Zhang_1997}. We assumed that the central source has a wind bound inside a cone with
an opening angle of $40^{\circ}$ along the polar axis. Table\, \ref{model:irs1w_2}
displays the fitted parameters and the third row of images in Figure\,
\ref{fig:models} represents the best-fits for the narrow solution. A comparison
between the isotropic wind model and the narrow solution (see
Fig. \ref{fig:models}) demonstrates that the latter fits the observed
emission of IRS\,1W better, with the main difference being in a smaller
R$_{0}$ and different inclination. The narrow solution models have
considerably lower residuals and result in best-fit parameters with
very low scatter between the different observing wavelengths.

Another plausible explanation for the
shape of IRS\,1W can be changes in the density of the
ISM. This condition is not rare in complex environments like the
NA where filaments and clumps of gas/dust are observed. For example,
\citet{Rauch_2013} determined the geometry and emission of the
asymmetric IRS\,8 bow shock using a shell solution for
a quadratic density gradient in the NA
ISM. Figure\,\ref{fig:irs1w_grad} displays our best-fit solution to IRS\,1W in
$L'$ for an ISM density gradient. The density gradient function has the form
$\rho=\rho_{0}(1+a_{1}x+a_{2}x^2)$, where $\rho_{0}$ is the mean density of the
ISM and $x$ the direction of the $\rho_{0}$
increment. In our model, the density increased from east to
west and the values of the coefficients $a_{1}$ and $a_{2}$  were fixed
to 2.0 and 5.0, respectively. 

This bow-shock gradient solution produced residuals on the order of
those of the narrow solution. Similar values were obtained between the
narrow and the gradient solution for the fitted
parameters. The best-fit gradient solution produced a $R_{0}$=533 AU, and
inclination of $i=$103$^{\circ}$ and a PA=9$^{\circ}$. Nevertheless, to perform a complete least-squares fitting to bow-shock
solutions with free gradient coefficients and/or different gradient functions
creates strong degeneracies among the fitted parameters. Hence, we
did not investigate other gradient solutions and
decided to use the narrow solution best-fit model for the
following analysis. Although we cannot fully distinguish between the
physical mechanisms that produce the observed shape of IRS\,1W, our
narrow and gradient models suggest that inhomogeneities in the
stellar wind and/or ISM are important in shaping the structure of this
particular bow shock.

\begin{table}[ht]
\caption{Best-fit parameters of our bow-shock models for IRS\,1W
  (spherically symmetric stellar wind) \label{tab:irs1w_1}} % title of Table
\centering  % used for centering table
\begin{tabular}{l c c c c } % centered columns (5 columns)
\hline
Parameter &$H$ &$K_s$ &$L'$ &$\sigma_{rms}$\\  
\hline                  % inserts single horizontal line
\hline
$R_{0}$ [AU] & 689.0 & 606.0 & 651.0 & $\pm34.0$\\ % inserting body of the table
P.A. [deg] & 3.0 & -13.0 & 27.0 & $\pm16.0$\\ 
Inclination [deg] & 131.0 & 121.0 & 127.0 & $\pm4.0$\\ 
FWHM [px] & 2.0 & 3.24 & 4.6 & - \\
Bow-shock flux [\%] & 49.0 & 79.0 & 100.0 & - \\ 
\hline
\end{tabular}
\end{table} 

\begin{table}[ht]
\caption{Best-fit parameters of our bow-shock models for IRS\,1w
  (narrow solution) \label{model:irs1w_2}} % title of Table
\centering  % used for centering table
\begin{tabular}{l c c c c } % centered columns (5 columns)
\hline
Parameter &$H$ &$K_s$ &$L'$ &$\sigma_{rms}$\\  % inserts table 
\hline                  % inserts single horizontal line
\hline
$R_{0}$ [AU] & 520.0 & 523.0 & 523.0 & $\pm1.2$\\ % inserting body of the table
P.A. [deg] & 13.0 & 2.0 & 5.0 & $\pm5.0$\\ 
Inclination [deg] & 88.0 & 96.0 & 84.0 & $\pm5.0$\\ 
FWHM [px] & 2.0 & 3.42 & 4.6 & - \\
Bow-shock flux [\%] & 50.0 & 79.0 & 100.0 & - \\ 
\hline
\end{tabular}
\end{table}

\begin{figure}
\centering
\includegraphics[width=\columnwidth]{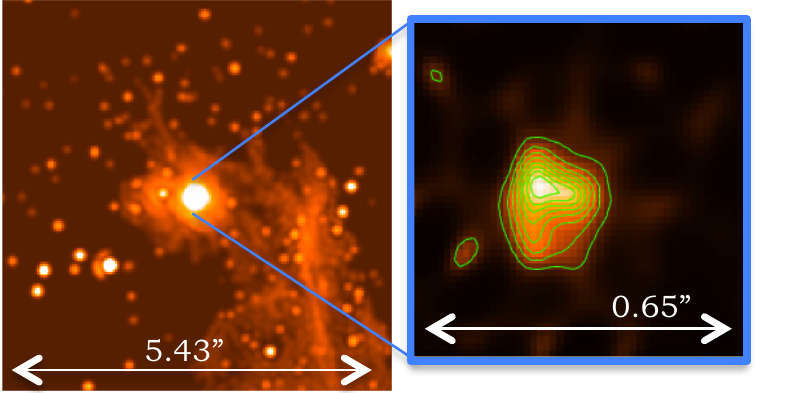} 
\caption{\textbf{Left:} $L'$-band image of the surroundings of IRS\,10W. The
  filamentary structure of the NA is appreciated as well as some
  additional close-by stars with extended emission. \textbf{Right:}
  the reconstructed SAM image of
  IRS\,10W in $L'$ band. An asymmetric structure can be seen. Contours represents 10, 20, 30, 40, 50, 60, 70, 80, and
  90 \% of the peak flux.}
 \label{fig:irs10w_env}
\end{figure}

\begin{figure*}[htp]
\centering
\includegraphics[width=\textwidth]{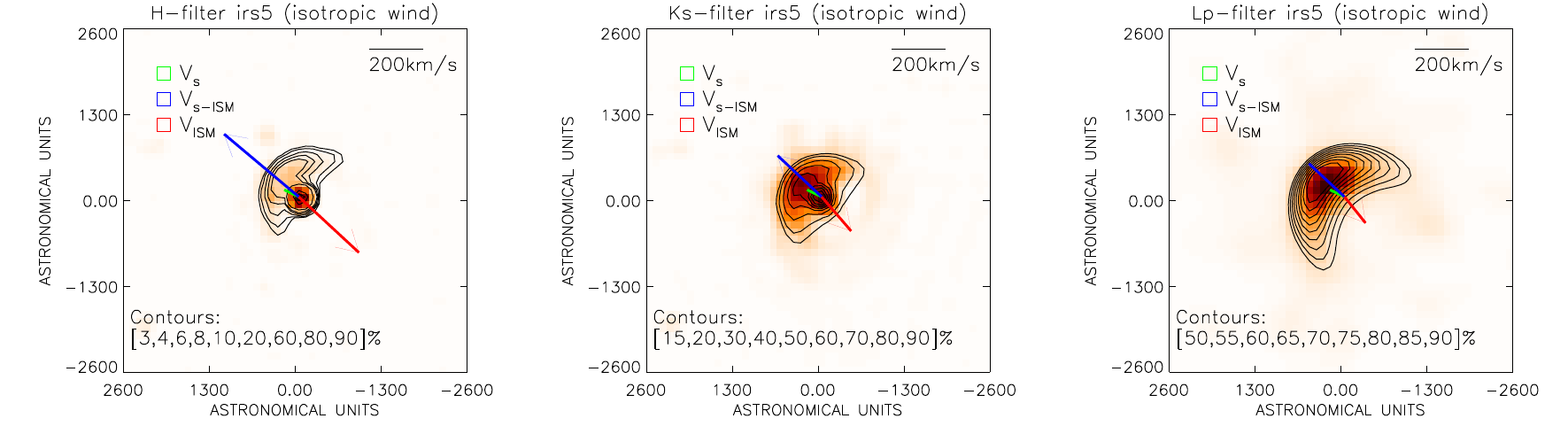} 
\includegraphics[width=\textwidth]{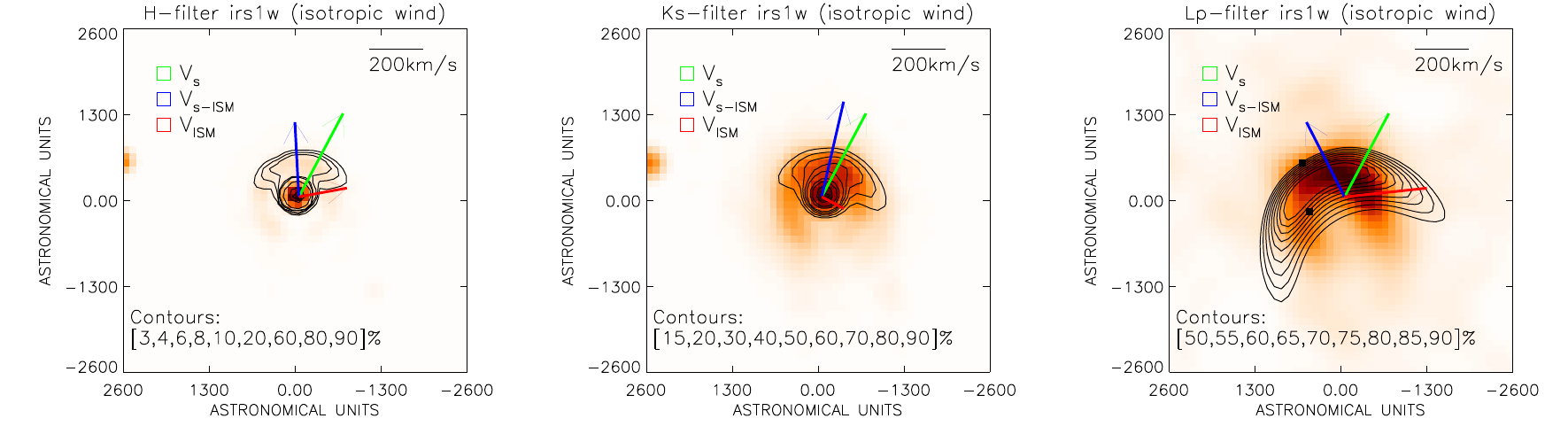} 
\includegraphics[width=\textwidth]{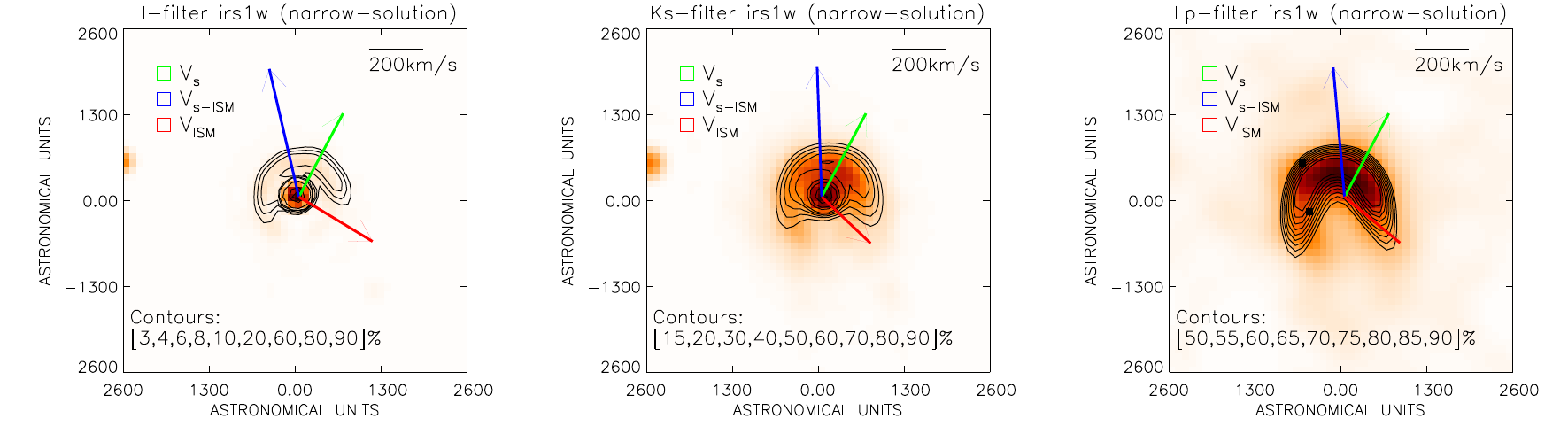} 
\includegraphics[width=\textwidth]{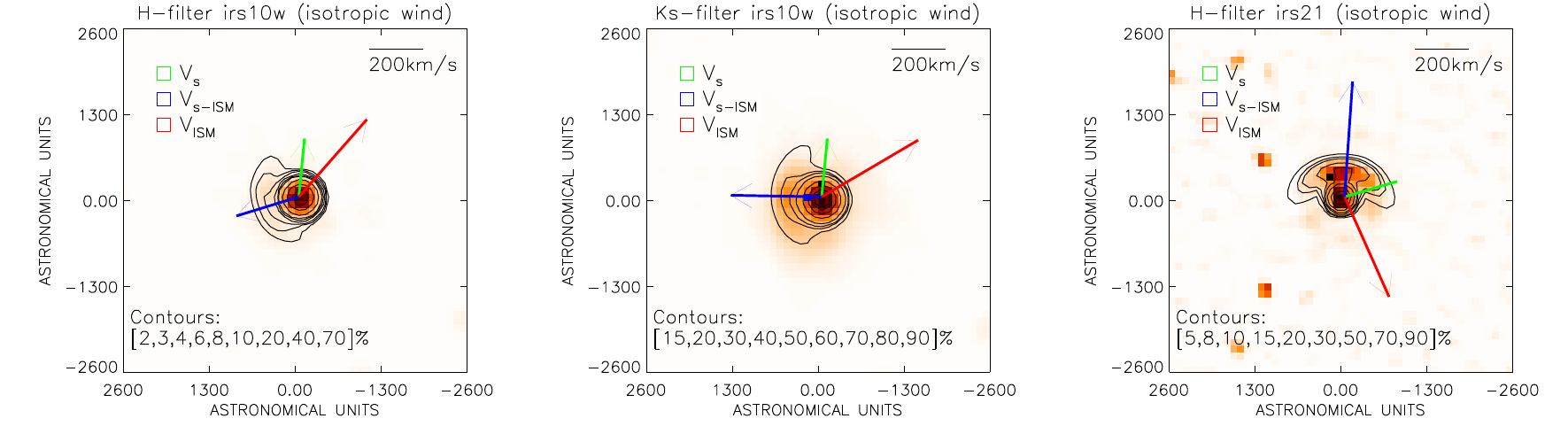} 
\caption{Bow-shock fitting of IRS 1W, IRS 5, IRS 10W and
    IRS 21 in H, $K_s$ and L'
    wavelengths. We represent in colors the real image and in
    contours the best model obtained. The colored vectors represent
    the velocity of the star ($V_s$; green arrow), the relative motion of the star
    through the ISM ($V_{s-ISM}$; blue arrow) and the dust motion of the ISM
    obtained with our bow-shock model ($V_{ISM}$; red arrow). \label{fig:models}}
\end{figure*}

{\bf IRS\,10W:} the extended emission around this source presents a
particularly different morphology in $L'$ , more so than in $H$ and $K_s$. The
central star is detected in the $H$ and $K_s$ filters. Extended
emission is observed at the position of the central source in $L'$. It
is unclear whether this emission is related to the central star or
not. The projected
emission in $H$ and $K_s$ displays a
symmetric structure, while in $L'$ the southward pointing tail is more
extended than the northward pointing one. This feature may be related
to inhomogeneities in the NA ISM at the location of IRS\,10W (see
Fig.\, \ref{fig:irs10w_env}). The radio image mapped at 1.3 cm by
\citet[][see their Fig. 10]{Zhao:2009la}
suggests that IRS\,10W is part of the NA  and is surrounded by ionized gas in a shell-like
structure with a completely different morphology than the one
observed at NIR wavelengths, showing no clear evidence of a bow-shock. 

Because of the asymmetric shape of IRS\,10W
in  $L'$, we decided to apply the model fitting only to the $H$ and
$K_s$ images. Table\, \ref{model:irs10w} shows the
best-fit parameters. The fourth row of images in
Figure\, \ref{fig:models} displays the best-fit models in contours. Although
the extended emission around IRS\,10W can be interpreted as a bow shock, a more detailed analysis of the IRS\,10W proper motion
and of the NA motion (at the position of IRS\,10W) does not support the idea of a
bow shock produced by the relative motion of the star throughout the
ISM. 

IRS\,10W has a proper motion of  217$\pm$14 kms$^{-1}$ with a PA of
355$^{\circ}$ E of N (see
Fig\,\ref{fig:models} and Table\,\ref{tab:proper_motion}). \citet{Paumard:2004kx} reported a gas flow of
189$\pm$84 kms$^{-1}$ with a PA of 188$^{\circ}$ E of N 
at the position of IRS\,10W, while the values according to \citet{Zhao:2009la} are 229$\pm$49 kms$^{-1}$ with a PA of
191$^{\circ}$. Therefore, the relative proper motion of 
IRS\,10W with respect to the NA is 403$\pm$86 kms$^{-1}$ at a PA of 1$^{\circ}$
E of N, or 441$\pm$37 kms$^{-1}$ at a PA of 3$^{\circ}$
E of N, that is, at a PA perpendicular to the direction of the possible
bow-shock symmetry axis (around $\sim$90$^{\circ}$). This suggests that the extended
emission around  IRS\,10W is not produced by the bow-shock model
described in Sect. 4.2, which implies that the relative motion and
the bow-shock axis must be parallel. Alternative explanations
for the observed
extended emission are (i) a shock between the stellar wind with a dust knot
or filament of the NA, (ii) a shock between the stellar winds of IRS\,10W and
a secondary source, or (iii) a chance superposition of a some NA feature
with IRS\,10W. These hypotheses are plausible since IRS\,10W is located
in a very mixed environment of the extended emission in the NA, some filaments can be observed superimposed along the source
position and, additionally, some close-by stars have
extended emission (see Fig.\, \ref{fig:irs10w_env}).

\begin{table}[ht]
\caption{Best-fit parameters of our bow-shock models for IRS\,10W\label{model:irs10w}} % title of Table
\centering  % used for centering table
\begin{tabular}{l c c c c } % centered columns (5 columns)
%\hline \hline                        %inserts double horizontal lines
%\multicolumn{5}{c}{Fitted Wavelengths} \\
%\hline \hline                        %inserts double horizontal lines
\hline
Parameter &$H$ &$K_s$ &$L' $ &$\sigma_{rms}$\\  % inserts table 
%heading
\hline                  % inserts single horizontal line
\hline
$R_{0}$ [AU] & 620.0 & 525.0 & - & $\pm48.0$\\ % inserting body of the table
P.A. [deg] & 107.0 & 98.0 & - & $\pm9.0$\\ 
Inclination [deg] & 123.0 & 89.0 & - & $\pm13.0$\\ 
FWHM [px] & 2.88 & 3.60 & - & - \\
Bow-shock flux [\%] & 26.0 & 45.0 & - & - \\ 
\hline
\end{tabular}
\end{table}

{\bf IRS 21:} the bow-shock of this source can be clearly seen at all
wavelengths and is similar to the deconvolved $K_s$-band image
presented by \citet{Buchholz:2011fk}. IRS\,21 is best resolved in H
band, where the central star is also unambiguously detected. We can
discard the hypothesis, made by \citet{Tanner:2005fk}, that IRS\,21 is seen face-on, since these authors could not resolve its structure
in their lower Strehl no-AO observations. In our $K_s$ and $L'$ images the central
source is not detected and the bow-shock morphology is less clearly
observed because of the lack of angular resolution. We therefore decided to apply the model fitting only to the
$H$-band image. The steady-state model fits the morphology
observed in the $H-$band very well. The proper motions of the star and of the NA environment are consistent
with the relative motion of the bow shock obtained from our
model. Table\, \ref{model:irs21} lists the best-fit parameters.

Since we only performed the model fitting to the IRS\,21 $H-$band image,
the uncertainties in the fitted parameters were estimated in a different
way than for the other bow shocks, for which we have at least two
models in two different filters. For IRS\,21, the uncertainties
reported in Table\,\ref{model:irs21} were estimated by testing three
different starting values for the bow-shock inclination angle. We
chose this approach because the inclination angle strongly affects the determination of the
bow-shock stand-off distance. In fact, it is degenerated for low- and
high-inclinations and therefore, our code works best at intermediate values of 
inclination angles ($40^{\circ} \le i \le160^{\circ}$). Very low- or
high-inclination angles result in bow-shock shapes that resemble
spherical shells. Thereupon, our code cannot compute the
  stand-off distance since it strongly depends on the inclination
  angle (see
Section\,\ref{bs_fitting}). For IRS\,21 we tested the code
with initial conditions of $i=$ 50$^{\circ}$, 90$^{\circ}$,
145$^{\circ}$. The fitted parameters are reported in
Table\,\ref{model:irs21} and their errors are the standard deviation
of the three trials. Although the model did not always converge to the same solution, which can be largely explained by degeneracies between the model parameters, the best-fit parameters, resulting from the runs with significantly different starting values for the inclination, have overall similar best-fit parameters with a low scatter between them.

\begin{table}[ht]
\caption{Best-fit parameters of our bow-shock models for IRS\,21
  (steady-state) \label{model:irs21}} % title of Table
\centering  % used for centering table
\begin{tabular}{l c c c c } % centered columns (5 columns)
%\hline \hline                        %inserts double horizontal lines
%\multicolumn{5}{c}{Fitted Wavelengths} \\
%\hline \hline                        %inserts double horizontal lines
\hline
Parameter &$H_{i=50^{\circ}}$ &$H_{i=90^{\circ}}$ &$H_{i=145^{\circ}}$  & $\sigma_{rms}$\\ % inserts table 
%heading
\hline                  % inserts single horizontal line
\hline
$R_{0}$ [AU] & 432.0 & 417.0 & 462.0 & $\pm$22.0\\ % inserting body of the table
P.A. [deg] & -8.5 & -4.0 & -6.3 & $\pm$2.2\\ 
Inclination [deg] & 106.0 & 95 & 116 & $\pm$10.0\\ 
FWHM [px] & 2.14 & 2.11 & 2.05 &  $\pm$0.04\\ 
Bow-shock flux [\%] & 68.0 & 62.0 & 65.0 & $\pm$3.0 \\       % [1ex] adds vertical space
\hline
\end{tabular}
\end{table}

\section{ISM velocity field and bow-shock orbital parameters \label{sec:Orbital_param}}

\subsection{Northern Arm dust motion \label{subsec:dust_motion}} 

Our best-fit bow-shock models provide estimates of the local
motion of the ISM in the NA. Assuming that our objects are
Wolf-Rayet stars \citep{Tanner:2005fk} with a mass-loss rate on the
order of  $\dot{m_{*}} \sim $1-2$\times$10$^{-5}$M$_{\odot}/$yr
and stellar winds with velocities around $\nu_{w} \sim$1000 km
s$^{-1}$, and that the density in the NA has a mean particle
number $n_{H} \sim$3$\times$10$^4$
cm$^{-3}$ \citep[see][]{Tanner:2002kx, Rauch_2013}, we  first calculated the relative velocity between star
and ISM with Eq.\,\ref{eq:sd}. Subsequently, we d the measured proper
motion of the central star to calculate the proper motion of the ISM at the
position of the bow-shocks. Figure\, \ref{fig:models} shows the projected motion of
the ISM, $V_{\rm ISM}$(red
vector), the relative motion of the bow shock, $V_{s- \rm ISM}$, according
to our best-fit
model (blue vector), and the proper motion, $V_{s}$, of the central source
(green vector) for all our targets. All our ISM motion measurements (magnitude and
position angle) agree to within 1-$\sigma$  with the models of
\citet{Paumard:2004kx} and/or \citet{Zhao:2009la} except for
IRS\,10W, which supports our hypothesis that IRS\,10W does not
represent a standard bow shock (see section 4.5). Table\,\ref{tab:NA_motions} summarizes the
NA projected motions obtained in this work and the previously published models of gas
and dust.  These results corroborate the hypothesis of
Wolf-Rayet stars as the central sources of the bow shocks. For
comparison, if we compute the projected NA
motions using O-stars, \citep[$\dot{m_{*}}
\sim$10$^{-7}$M$_{\odot}/$yr ; $\nu_{w} \sim$2000 km
s$^{-1}$;][]{Prinja_1990} as
the central engines of the bow-shocks, the resulting NA motions are
a factor of $\sim$5-10 lower than the observationally inferred values. Hence, we discard the idea
of O-stars as the central sources of the bright NA
bow shocks.

\begin{table*}[ht]
\caption{ Velocities of the NA ISM considering the bow-shock model
  used in this work, the \citet{Paumard:2004kx}, and the
  \citet{Zhao:2009la} models. \label{tab:NA_motions}} % title of Table
\centering  % used for centering table
\begin{tabular}{l c c c c c c || c c c  } % centered columns (5 columns)
\hline \hline 
\multicolumn{7}{c}{ Bow-shock model NA velocities [km s$^{-1}$]} & \multicolumn{3}{c}{Position angles [deg.]} \\
\hline 
Source & $\nu_{\rm ISM-R.A.}^{\mathrm{a}}$ & $\nu_{\rm ISM-Dec.}^{\mathrm{a}}$
& $\nu_{\rm ISM-R.A.,P.}^{\mathrm{b}}$ &
$\nu_{\rm ISM-Dec.,P.}^{\mathrm{b}}$ &
$\nu_{\rm ISM-R.A.,Z.}^{\mathrm{c}}$ & $\nu_{\rm
  ISM-Dec.,Z.}^{\mathrm{c}}$ & P.A.$_{\rm ISM}^{\mathrm{a}}$ &
P.A.$_{\rm ISM,P}^{\mathrm{b}}$  & P.A.$_{\rm ISM,Z}^{\mathrm{c}}$ \\
\hline 
IRS\,5  & -136$\pm$76 & -143$\pm$56 & -0.88$\pm$67 &
-133$\pm$91 &-54 $\pm$20 & -141$\pm$13 &  223$\pm$33 & 0.4$\pm$207 & 201$\pm$8 \\
IRS\,1W & -220$\pm$48 & -169$\pm$4 & -73$\pm$49 & -213$\pm$69 &
-218$\pm$13 & -110$\pm$13 & 232$\pm$16 & 199$\pm$19 & 243$\pm$15\\
IRS\,10W & -305$\pm$74 & 248$\pm$54 & -28.54$\pm$63 & -187$\pm$85  &
-44$\pm$29 & -225$\pm$50 & 310$\pm$76 & 189$\pm$19 & 191$\pm$11 \\
IRS\,21 & -123$\pm$16 & -331$\pm$41 & -127$\pm$54 & -325$\pm$51 & -275$\pm$28 &
-250$\pm$66 & 200$\pm$13 & 201$\pm$17 & 228$\pm$80 \\
\hline
\end{tabular}
\begin{list}{}{}
\item[$^{\mathrm{a}}$] Motion of the ISM and its position angle measured in this work.
\item[$^{\mathrm{b}}$] Motion of the ISM and its position angle measured by \citet{Paumard:2004kx}.
\item[$^{\mathrm{b}}$] Motion of the ISM and its position angle measured by \citet{Zhao:2009la}.
\end{list}
\end{table*}

\subsection{Orbital motion of the bow-shock sources}

The location of the sources that we studied within the NA and the
determination of their geometry with the models used here provide valuable information to determine the probability density distribution
(PDF) of their orbital parameters,  $PDF (a,e,i,\Omega,
\omega, P, T_0)$. We used the methodology described by
\citet{Lu:2009bl} and  the same definitions for the orbital elements. This method consists of a Monte Carlo simulation
of $10^4$ trials to convert the 3D positions and velocities of the
sources into their six orbital parameters. The projected positions of the
bow shocks are obtained from the astrometric measurements used to
compute their proper motions (see Sect. 3). The distances of the bow shocks along the LOS and their
respective uncertainties were obtained from the NA model of
\citet{Paumard:2004kx}. The LOS velocities were constrained in two
different ways. On the one hand, they were constrained by assuming that
the sources are tied to the gravitational potential of SgrA* and
have a 3D motion lower than their escape velocities. This method provides an upper limit for
the magnitude of the LOS velocity but with an ambiguity in its sign. The PDFs of the orbital parameters were therefore constructed using both signs for the LOS velocities. The second method
computes the LOS velocity using
the relative velocity of the sources through the ISM obtained from our model fitting (see Sects.
\ref{subsec:shape}, \ref{subsec:dust_motion}). With this method we
determined an allowed 3-$\sigma$ range of LOS velocities. Furthermore,
this approximation helped us eliminate the
degeneracy of sign presented in the derivation of the LOS velocity
using the first method.   
The computed escape velocities, the 3D positions, the proper motions,
and the 3-$\sigma$ ranges for the LOS velocities are reported in
Table\,\ref{tab:BS-velocities}. The 3D positions and proper motions
use a Gaussian distribution as prior, while the LOS velocity uses a
uniform distribution. These
parameters allowed us to determine the $PDF (a,e,i,\Omega,
\omega, P, T_0)$ of each one of our sources.

After obtaining the PDFs of the orbital parameters, we computed
the orientation of the
sources' orbital planes in the sky, as viewed from SgrA*. The orientation of the orbital planes can be described by a
unitary vector ($\bar{s}$) originating at the position of Sgr\,A* and normal
to the source's
orbital plane. To determine the Cartesian components of $\bar{s}$, the
PDFs of the inclination angle, $i$,  and the ascending node, $\Omega$,
were used according to the following expression:

\begin{equation}
\bar{s}= 
\left( \begin{array}{c}
s_x \\
s_y \\
s_z
\end{array}
\right) = 
\left(\begin{array}{c}
sin\,i\,\,cos\,\Omega \\
-sin\,i\,\,sin\,\Omega \\
-cos\,i 
\end{array} 
\right)
.\end{equation}

\begin{figure*}
\centering
\includegraphics[width=18cm, height=5.8cm]{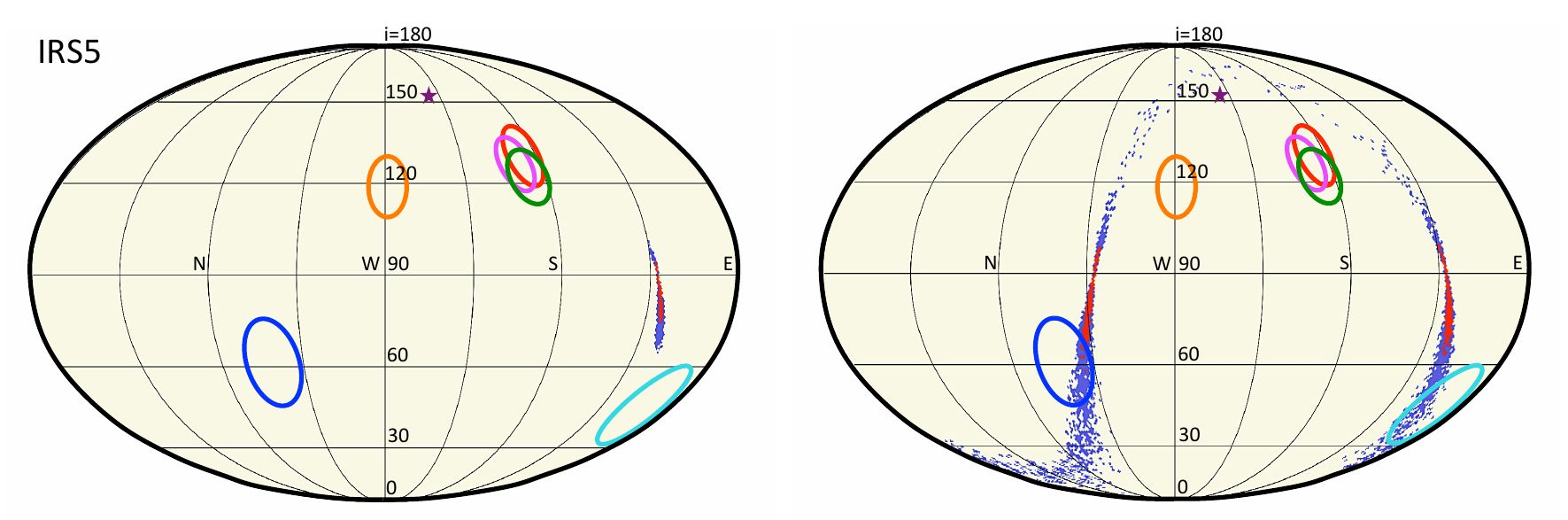} 
\includegraphics[width=18cm, height=5.8cm]{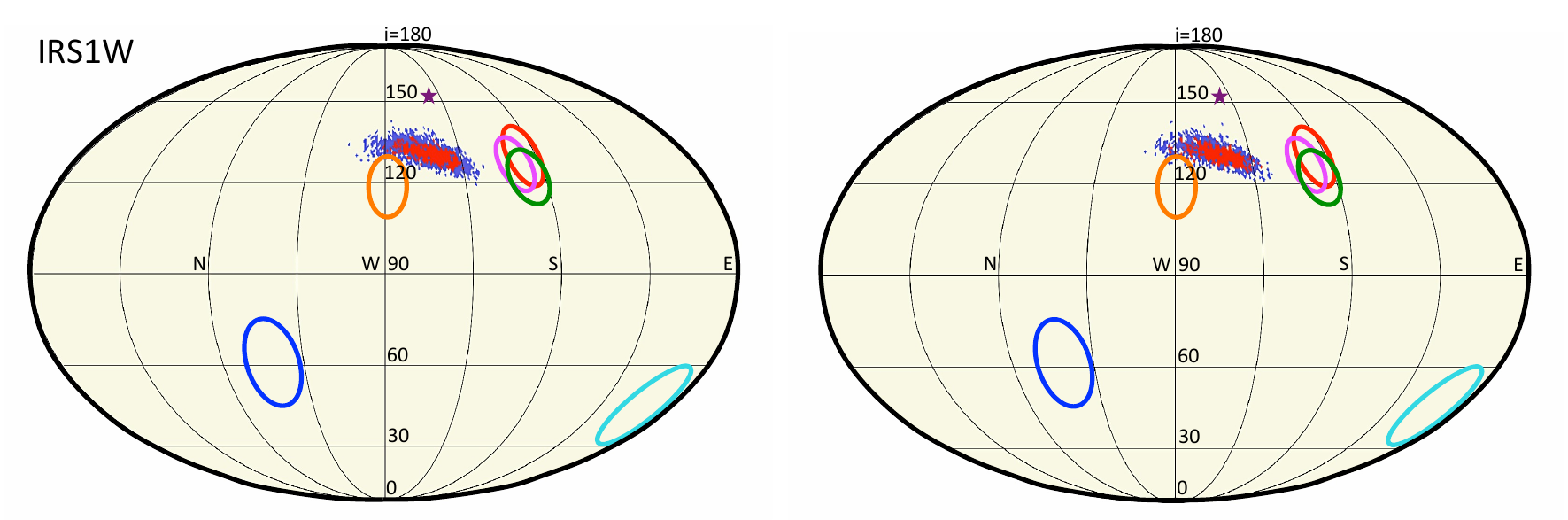} 
\includegraphics[width=18cm, height=5.8cm]{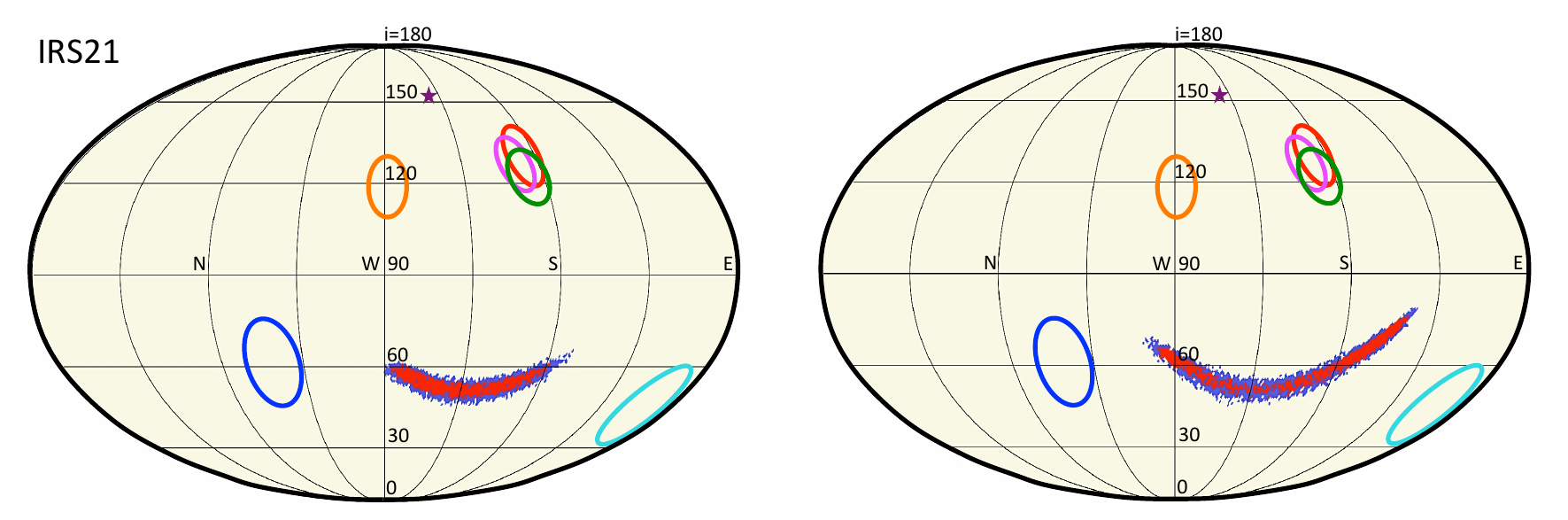} 
\includegraphics[width=18cm, height=5.8cm]{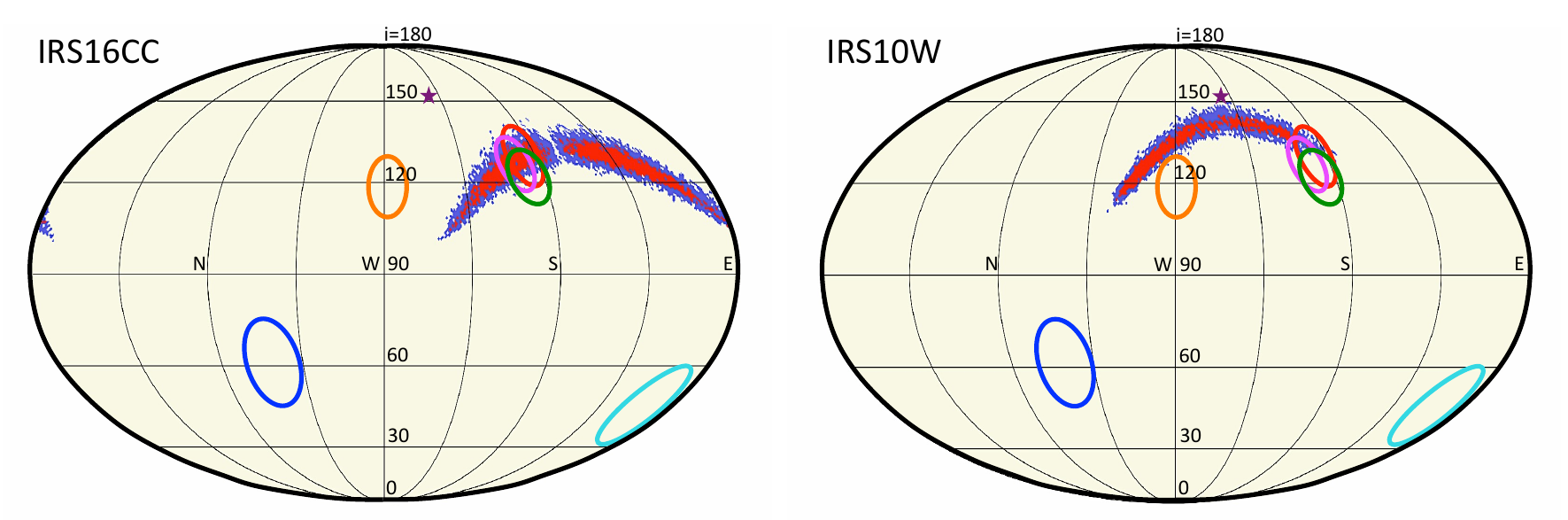} %width=\textwidth 
\caption{Distribution of the orbital planes of IRS\,5, IRS\,1W,
  IRS\,21, IRS\,10W, and IRS\,16CC. The 1-$\sigma$ and 2-$\sigma$
  confidence levels are shown in red and blue. The ellipses represent the 1-$\sigma$ position of the CCW, CCWS,
  and the NA orbital planes: the NA is represented by the light-blue
  ellipse \citep{Paumard:2006xd, Bartko:2009fq}; the green ellipse corresponds to the CWS
  according to \citet{Lu:2009bl}; the red, pink, and orange ellipses
  display the CWS according to \citet{Bartko:2009fq} for stars at
  0.8''-12'', 0.8''-3.5',' and 7''-12'' projected
  distances, respectively; the purple
  star represents the position of the CWS for stars at  3.5''-7''
  projected distance from Sgr\,A* \citep{Bartko:2009fq}, and  the blue
  ellipse displays the position of the CCWS according to \citet{Paumard:2006xd}  }
 \label{fig:map}
\end{figure*}

Figure\,\ref{fig:map} displays the pointing vectors obtained with the
PDFs of the orientation of the sources' orbital planes using a Hammer projection. The left column of the maps
shows the vectors of the PDF$(i, \Omega)$ of the bow shocks using the computed 3-$\sigma$ LOS
range. The right column shows the vectors of the PDF$(i, \Omega)$ using only the
3-$\sigma$ LOS upper limit determined from the escape velocity. For IRS\,10W we only computed its probability distribution using the
3-$\sigma$ LOS upper limit since this source was not clearly
identified as a bow shock. Additionally, we included the PDF$(i, \Omega)$ of
IRS\,16CC, a source previously studied by \citet{Lu:2009bl}, to check the validity of our code. We obtain similar results to those previously published, which confirms
the accuracy of our code. Each map in Fig.\,\ref{fig:map} displays
the 1-$\sigma$ and 2-$\sigma$ confidence levels of the PDFs in red and
blue. The confidence levels were computed using only orbital
solutions tied to Sgr\,A*. The ellipses plotted in different colors represent the 1-$\sigma$ contour of the orbital
planes of some of the main dynamical structures at the GC. Table\,\ref{tab:BS_orb_planes} shows
the peak values of the PDFs for
the sources identified as bow shocks.    

The plots in the left and right columns of
Fig.\,\ref{fig:map} clearly show that the  PDFs  based on
the modeling of the bow-shocks
constrain the orbital planes significantly better than those
obtained with the constraints from the escape velocities alone. The
PDF($i, \Omega$) of IRS\,5 is the best constrained. The derived LOS
velocity range of IRS\,5 fully eliminates the
ambiguity in the sign of the LOS velocity. Conversely, IRS\,1W
exhibits very similar PDFs using both methods. This effect can be explained by the
fact that we only took into account orbits gravitationally bound to
Sgr\,A*. Furthermore, most of the IRS\,1W bound orbits obtained from our MC
simulation have LOS velocities
within the 3-$\sigma$ range derived from our bow-shock modeling. From
this analysis we also observe that IRS\,5
and IRS\,21 clearly are not part of the CWS or the CCWS. IRS\,1W
and IRS\,10W are relatively close to the position of the CWS but still on
the edge of the CWS
1-$\sigma$ limits. Hence, we cannot establish with certainty whether they are CWS
members or not. None of the sources are moving in the same plane of the NA. 

\begin{table*}[ht]
\caption{ 3D coordinates and velocities of IRS\,1W, IRS\,10W, IRS\,21,
  and IRS\,5.\label{tab:BS-velocities} } % title of Table
\centering  % used for centering table
\begin{tabular}{r c c c c } % centered columns (5 columns)
\hline\hline                        %inserts double horizontal lines
Star &IRS\,1W &IRS 5 &IRS 10W &IRS 21\\ % inserts table 
%heading
\hline                  % inserts single horizontal line
$x\quad[arcsec]^{\mathrm{a}}$& 5.26$\pm$0.009 & 8.62$\pm$0.02 &
6.5$\pm$0.01 & -2.35$\pm$0.001\\ 
$y\quad[arcsec]^{\mathrm{a}}$ & 0.616$\pm$0.016 &9.83$\pm$0.032 &
5.15$\pm$0.02 & -2.69$\pm$0.001\\ 
$z\quad[arcsec]^{\mathrm{b}}$ & 7.53$\pm$1.53 &3.81$\pm$2.66 &
6.43$\pm$1.5 & 4.45$\pm$1.07\\ 
$v_{e} \quad[km/s]^{\mathrm{c}}$ & 320$\pm$29 &266$\pm$7 & 303$\pm$13 &
412$\pm$30\\ 
$v_{R.A.}\quad[km/s]^{\mathrm{d}}$ & -164$\pm$35 & 54$\pm$18 & -21$\pm$21 &
165$\pm$13 \\ 
$v_{Dec.}\quad[km/s]^{\mathrm{d}}$ & 309$\pm$22 & 26$\pm$25 & 216$\pm$14 & 57$\pm$ 33\\
$v_{LOS}\quad[km/s]^{\mathrm{e}}$ & [-18, 18]  & [113, 334] &
[-307, 307] & [-79, 261] \\ 
\hline
\end{tabular}
\begin{list}{}{}
\footnotesize{\item[$^{\mathrm{a}}$] Projected distances  from SgrA*.}
\footnotesize{\item[$^{\mathrm{b}}$] LOS distance from the plane containing SgrA*, using the NA model of \citet{Paumard:2004kx}}.
\footnotesize{\item[$^{\mathrm{c}}$] 3D escape velocity at the
  position of the source.}
\footnotesize{\item[$^{\mathrm{d}}$] Proper motion measured in this work.}
\footnotesize{\item[$^{\mathrm{e}}$] 3-$\sigma$ range on the LOS velocity of the bow shocks.}
\end{list}
\end{table*}

\section{Discussion and conclusions \label{sec:discussion}}

The AO facilities and the sparse aperture masking mode of NACO at the
ESO VLT allowed us to perform a NIR multiwavelength campaign to
study with high-angular resolution the structure, circumstellar matter, and
kinematics of some of the brightest and most embedded sources in the
central parsec of the GC. We used AO-deconvolved and beam-restored
images in $H$ and $K_s$ and reconstructed SAM images in $L'$ to
determine the morphology of IRS\,5, IRS\,1W, IRS\,10W, and
IRS\,21. We presented a new technique for calibrating SAM data of
dense star fields that does not require observations of an off-target
calibrator source. Our new procedure, called the synthesized calibrator, uses our a priori information of the stars in the field
and constructs a median superposition of bona-fide point-like sources
to build the final calibrator. This method allowed us to minimize
the noise, poor background estimates, and problems in the PSF
stability. It is also an alternative way of calibrating SAM observations where traditional
calibrators are too dim or located far away from the science target,
like in the case of the GC. Additionally, we used AO observations in
$K_s$ between 2002 to 2011 to derive the proper motions of our
sources. 

We interpreted our observations with 3D steady-state bow-shock models
and confirmed that three of them (IRS\,5, IRS\,1W, and IRS\,21) are bow shocks
associated with the relative motion of Wolf-Rayet stars
through the ISM. Although this conclusion was suggested previously,
our models give the best constraints on the bow-shock shell
and kinematics so far because of our detailed modeling, new proper motion
data, and the inclusion of new NA proper motion data from
\citet{Zhao:2009la}, which allowed us an additional consistency
check. In contrast to \citet{Tanner:2005fk}, we also found a good model
fit to the IRS\,1W. On
the other hand, the extended emission around IRS\,10W does not appear to trace a bow shock, but may be produced by the shock of two stellar
winds, or there may be a chance superposition with a mini-spiral
feature, or the star may heat a nearby mini-spiral filament. We also confirmed that IRS\,21 is an edge-on bow shock,
not a face-on bow shock, as was previously hypothesized by
\citet{Tanner:2005fk}. 

Our results support the idea
that the observed NIR bow-shock emission is thermal emission from hot
dust grains. \citet{Tanner:2002kx} fit a 1000 K black-body
to the spectral energy distribution of IRS\,21. Dust can easily reach
such temperatures in the local environs of massive (O/WR)
stars since they are strong UV radiation emitters. For
our bow-shock sources, which have IR luminosities of L$\sim$10$^4$-10$^5$
L$_{\odot}$\citep{Tanner:2005fk}, the smallest distance from the
central source at which dust can exist without evaporation, $r_{evap}$,
can be calculated, at first glance, as \citep{Barvainis_1987} 

\begin{equation}
r_{evap}=1.3\sqrt{L_{UV,39}}\,T_{1500}^{-2.8}\,pc
,\end{equation}

where $L_{UV,39}$ is in units of 10$^{39}$ watts and $T_{1500}$
in units of 1500 K. Hence, $r_{evap}$ for our sources is
$\sim$8$\times$10$^{-5}$-25$\times$10$^{-5}$ pc or 16-52 AU. This lies well inside
the measured size of the bow-shock shells of our targets at scales
lower than our angular resolution. Taking into account the grain 
sizes used in our study, we can also compute $r_{evap}$
using Eq.\,\ref{eq4}. In this case, the smallest distance at which dust can exist is 
about 24-75 AU for the given luminosities. Both approximations are
consistent, also with the observed bow-shock
sizes. From Eq.\,\ref{eq4} we obtain a rough estimate of 470 K for the temperature near the apices of the bow shocks. These calculations, together with our scattering
and thermal-emission modeling, support the assumption
that the radiation in the GC bow-shock sources arises from thermal
emission in dust grains. Therefore, the mechanism responsible for the observed polarization is the alignment of
asymmetric/elongated dust grains. The alignment mechanism implies a compression of 
the magnetic field of the Northern Arm in the environs of the
bow shocks and the increment of the local 
field density. These changes produce a rapid magnetic alignment of the
grains \citep{Buchholz_2013}. From the best-fit bow-shock stand-off
distances, we have established that their
central sources are massive objects (WR stars) with supersonic stellar
wind velocities and strong mass-loss rates. They are therefore
short-lived objects and have probably formed during the most recent
starburst event at the GC. 

One of the main sources of uncertainty in determining the orbital
planes of the population of massive stars around the GC is the lack of
information of the stellar LOS distances \citep[see
e.g.,][]{Lu:2009bl, Yelda_2014}. Therefore, their LOS distances
are significantly better constrained by the result that the extended
emission of our sources are bow-shocks produced by the interaction of
the stellar source with the NA. This characteristic, unique in this sources, along
with their LOS velocity from our best-fit models, their proper motions
and models of the ISM in the NA, allowed us to perform a Monte Carlo analysis of the sources' orbital
planes. 

IRS\,1W stands out from the other sources because of its
unusual morphology (see Sect. \ref{sec:results}), in particular its long tail, which we expect to trace the relative motion between IRS 1W and NA in the past 50 years. From studying the orbital-plane, we derived
that the most probable orbit for this source peaks at an
inclination of 132$^{\circ}$ with an ascending node angle of 154$^{\circ}$. Since
the NA orbital plane has an inclination of 45$^{\circ}$ with an
ascending node of 15$^{\circ}$, we discovered that both the IRS\,1W
and the NA orbital planes are approximately perpendicular to each other. Therefore,
 IRS\,1W is crossing the NA and its tail traces the thickness of the NA
 stream. Since the projected tail has an extension of 6400 AU, the NA thickness probably is on the order of $\sim$ 9000 AU at the
 position of this source.

The derived orbital planes suggest that our sources are not part of any of the suggested disks of young stars in the GC, but
our results are consistent with the most recent detailed
work on the orbital distribution of the young stars in the GC by
\citet{Yelda_2014}. These authors suggested that only about 20\% of 116
O/WR stars, at a projected distances of 0.8'' to 12'' from Sgr\,A*,
are part of the CWS. Hence, if our bow shocks were members of the CWS
, the number of disk members would be increased by about
20\%. However, our results show that the targeted bow-shock sources probably do not form part of the CWS. Therefore, our results provide new evidence for the complexity of the
population of young massive stars in the central parsec of the
GC and highlight that the disk members are
only a small fraction of the young massive stars around
Sgr\,A*. We conclude that models that intend to explain the formation
of the young massive stars in the central parsec of the GC and/or
their dynamical evolution need to be
adjusted to explain the presence of a considerable number of massive
stars with apparently random distribution of their orbital planes,
which suggests that either not all stars formed in a disk, or that the disk is rapidly dissolving. 

\begin{table}[ht]
\caption{ Maxima of the PDFs orbital planes for the inclination (i) and angle of the
  line of nodes in the sky ($\Omega$). \label{tab:BS_orb_planes} } % title of Table
\centering  % used for centering table
\begin{tabular}{l c c } % centered columns (5 columns)
\hline\hline                        %inserts double horizontal lines
Source &i [deg]$^{\mathrm{a}}$ & $\Omega$ [deg]$^{\mathrm{a}}$ \\%&$\sigma$ [deg]$^{\mathrm{b}}$ \\ % inserts table 
\hline                  % inserts single horizontal line
IRS\,1W & 132 & 154 \\ %& 13 \\
IRS\,5 & 78 & 37 \\%& 7 \\
%IRS\,10W & 11 & 21 %& 7 \\
IRS\,21 & 51 & 95 \\ %& 35 \\
\hline
\end{tabular}
\begin{list}{}{}
\footnotesize{\item[$^{\mathrm{a}}$] Angle between the orbit and the
  plane of sky.}  
\footnotesize{\item[$^{\mathrm{b}}$] Angle between North and the line
  of ascending nodes, increasing East of North. } 
\end{list}
\end{table}

\begin{acknowledgements}
The authors thank the referee for his/her useful comments that improved this
work. J.S.B., R.S. and
A.A. acknowledge support by grants AYA2009-13036-CO2-01, AYA2010-17631 and
AYA2012-38491-CO2-02 of the Spanish Ministry of Economy and
Competitiveness (MINECO) cofunded with FEDER funds, and by grant P08-TIC-4075 of the Junta de
Andaluc\'ia. R.S. acknowledges support by the Ram\'on y Cajal program
of the Spanish Ministry of Economy and
Competitiveness. J.S.B. acknowledges support by the JAE-PreDoc program of the Spanish Consejo Superior de
Investigaciones Cient\'ificas (CSIC) and to the ESO studentship program. Part of this work was supported
  by the COST Action MP0905: Black Holes in a violent Universe. This
  work is based on observations made with ESO Telescopes at the La
  Silla Paranal Observatory under programs 071.B-0077(A),
  073.B-0084(A), 073.B-0085(E), 077.B-0014(A), 081.B-0648(A),
  179.B-0261(X), 183.B-0100(D), 183.B-0100(G), 183.B-0100(V), and
  085.D-0214(A). 
\end{acknowledgements}

\bibliographystyle{aa}
\bibliography{/Users/JSB/Documents/BS-17Apr2013}

\appendix \section{Proper motion measurements \label{sec:propplots}}

\begin{figure*}[ht]
\centering
\includegraphics[width=\textwidth]{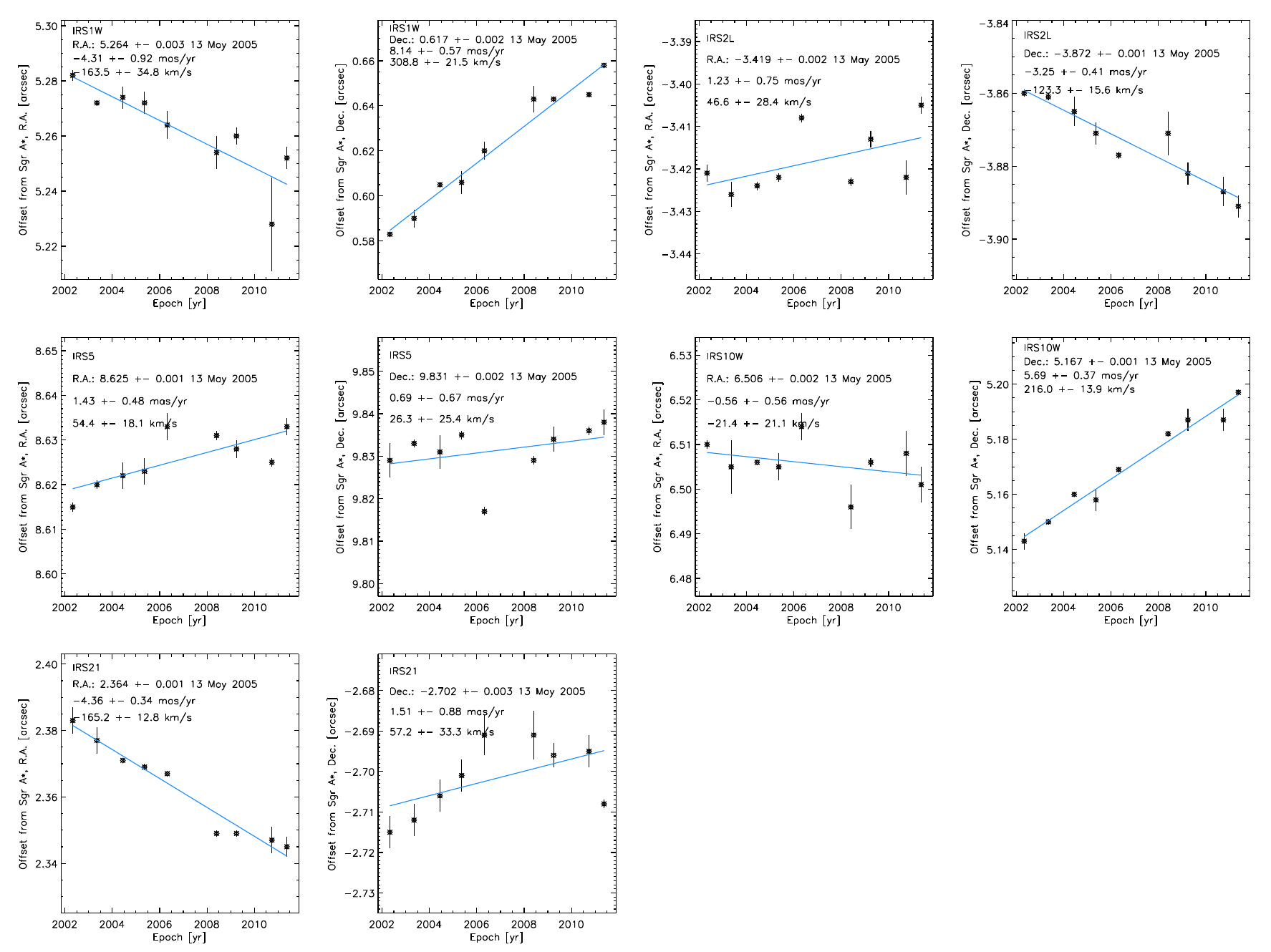} 
\caption{Proper motion measurements. Measured positions
  vs.\ time in right ascension and declination offsets from
  Sagittarius\,A* for the bow-shock sources IRS\,1W, IRS\,5, IRS\,2L, IRS\,10W,
  and IRS\,21.}
\end{figure*}

\begin{figure*}
\centering
\includegraphics[width=\textwidth]{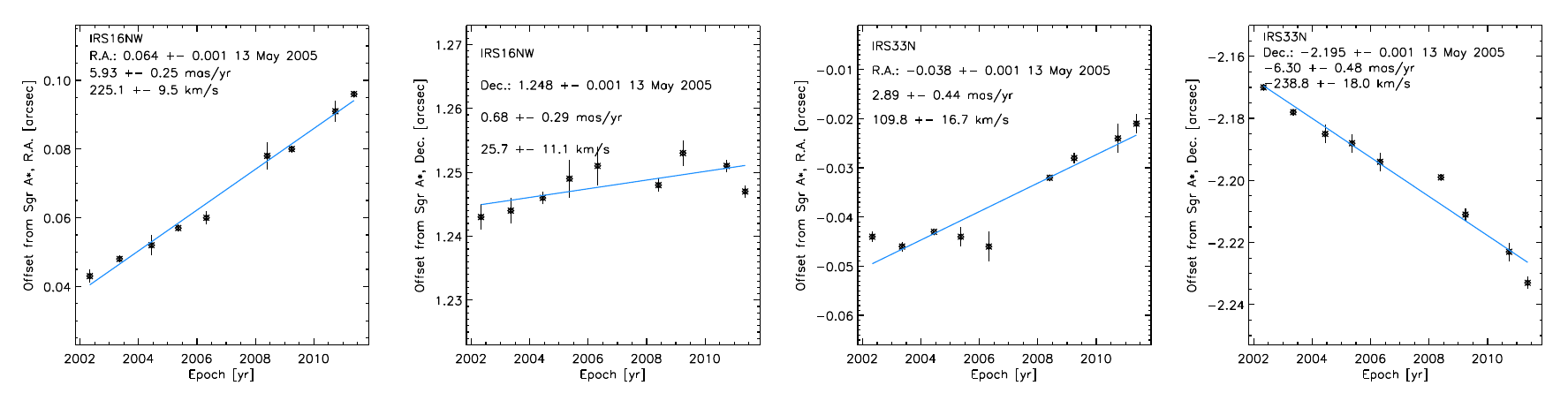} 
\caption{Proper motion measurements. Measured positions
  vs.\ time in right ascension and declination offsets from
  Sagittarius\,A* for the stars IRS\,16NW and IRS\,33N.}
\end{figure*}

\end{document}